\documentclass[fleqn,12pt,a4paper]{article}
\usepackage[a4paper, total={6in,9in} ]{geometry}


\usepackage{multirow}
\usepackage{theorem}
\usepackage{amsmath}
\usepackage{amssymb}
\usepackage{amsopn}
\usepackage{graphicx}
\usepackage[longnamesfirst]{natbib}
\usepackage[hidelinks]{hyperref}
\hypersetup{
	linkcolor=blue,
	filecolor=magenta,      
	urlcolor=blue,
	citecolor=blue,
}
\usepackage{xcolor}
\usepackage{subfigure}
\usepackage[font=small]{caption}
\usepackage{mathtools}

\usepackage[nameinlink]{cleveref}

\theoremstyle{break}
\theorembodyfont{\itshape}

\newtheorem{dfn}{Definition}

\begin{document}

\title{\textbf{Distilling Models of 
Bounded-Rational Choice: 
A Constraint Programming Approach 
}}

\author{\qquad \"{O}zg\"{u}r Akg\"{u}n 
	\qquad\qquad\qquad Georgios Gerasimou}
\date{\small July 2, 2026\thanks{
\"{O}zg\"{u}r Akg\"{u}n: School of Computer Science, University of St Andrews, 
\href{mailto:ozgur.akgun@st-andrews.ac.uk}{ozgur.akgun@st-andrews.ac.uk}. 
Georgios Gerasimou (corresponding author): 
Adam Smith Business School, University of Glasgow, 
\href{mailto:georgios.gerasimou@glasgow.ac.uk}{georgios.gerasimou@glasgow.ac.uk}.
Authors are ordered alphabetically. 
Georgios is grateful to Matúš Tejiščák for introducing him 
to \"{O}zg\"{u}r and to constraint programming.
We thank Paola Manzini for sharing with us the data that are analyzed 
in Section 5, and Thomas Demuynck for very helpful comments. 
The paper was presented at FUR 2026 (Alicante) and  
BSE 2026 (Barcelona), whose participants we thank for their feedback. 
Jacob Beadie, Tudor Huica and David Seruga 
provided excellent research assistance. 
Any errors are our own.
}}

\maketitle
\begin{abstract}  
We provide an analytical framework that allows 
for distilling the full explanatory 
and welfare-relevant content of influential yet   
computationally hard models of bounded-rational general choice.
We do so by introducing
constraint programming methods and tools 
from the optimization literature.  
We focus on the prominent ``shortlisting'' and 
``limited-attention'' models. 
Applying our framework on imperfectly rational human choice data,  
we find that these models jointly account for nearly all behaviors, 
with limited-attention ones explaining better while being 
more permissive. 
Selection criteria that we introduce narrow down the models' 
welfare-relevant predictions, 
considerably alleviating their indeterminacy and  
contributing toward their practical applicability. \\

\noindent Keywords: 
bounded rationality; 
general choice; 
constraint programming; 
shortlisting; 
limited attention;
predictive success.
\end{abstract}

\parskip=5pt

\setcounter{page}{0}

\thispagestyle{empty}

\vfill

\pagebreak

\section{Introduction}

Since the seminal work of \cite{kalai-rubinstein-spiegler02}, 
much of the development of bounded-rational choice theory in 
the past two decades has taken place within the analytical framework 
of deterministic general choice where the analyst is 
assumed to observe decision makers' behavior at  
finite menus of discrete 
alternatives.\footnote{This is presented in the  
introductory chapters of major graduate-level microeconomics 
textbooks such as 
\cite{kreps90}, \cite{mwg}, \cite{rubinstein06} and  
\cite{kreps12}.}
In addition to its generality, this environment's 
appealing features include the paucity of technical 
assumptions and interpretational clarity.
Because of these features, it provides a fertile ground 
for the development of theories of decision making that 
are more general than---or form alternatives to---the 
textbook model of standard utility maximization, 
or \textit{rational choice} (\textbf{RC}).

Arguably the two most influential theories of bounded-rational decisions 
in general choice environments 
are the intuitive 
\textit{``rational shortlist method''} by 
\cite{manzini-mariotti07} (henceforth \textbf{RS}, or \textit{``shortlisting''}) 
and the \textit{``choice with limited attention''} (\textbf{CLA}) procedure by 
\cite{masatlioglu-nakajima-ozbay12}, along with its 
\textit{``overwhelming choice''} (\textbf{OC}) 
extension by \cite{lleras-masatlioglu-nakajima-ozbay} 
(henceforth \textit{``limited-attention''} models).\footnote{
Other prominent contributions to this literature include  
\cite{masatlioglu-ok05,salant-rubinstein08, 
bernheim-rangel09}; and \cite{ok-ortoleva-riella15}.
A selection of additional contributions 
are discussed in the recent survey by \cite{declippel-rozen24}.
\cite{rubinstein_salant_hndbk} and several other 
chapters in \cite{caplin-shotter08} discuss methodological 
implications of the behavioral/bounded-rational  
approach to revealed preference.} 
The former portrays the decision maker as choosing  
in two stages: first, eliminating those feasible 
alternatives that are dominated according to some 
general shortlisting dominance relation; then, choosing  
from the set of shortlisted options by discarding  
those that are dominated according to a distinct second 
criterion. Limited-attention models 
are also two-stage choice procedures, but depict the agent 
as cognitively constrained in the sense of possibly being unable 
to consider \textit{all} feasible alternatives,  
and rational given this constraint: specifically, 
the agent maximizes a stable 
preference ordering over the alternatives that they actually 
consider at each menu. Both models include 
\textbf{RC}, as a special case: the former 
when the first dominance relation is a complete ordering;
the latter two when all feasible alternatives 
are always paid attention to.

While these seminal models have been studied at length in the 
theoretical literature, leading to many 
extensions, very few papers test their implications empirically 
and, to the best of our knowledge, none studies the particularly 
challenging task of doing so in the \textit{full generality} that is 
afforded by these models, or the problem of recovering the 
complete range of the models' behavioral primitives---by which we mean 
the underlying relations (preferences, attention filters)---in 
order to empirically assess their welfare-relevant information. 
This paper aims to facilitate such inquiries by 
introducing into the bounded-rationality 
and revealed-preference 
literatures the \textit{Constraint (Dominance) Programming} (C(D)P)
computational paradigms that achieve 
both these tasks simultaneously and, as highlighted above, in the 
most general formulations of the \textbf{RS},
\textbf{CLA} and \textbf{OC} models.
In addition, to alleviate the explanatory indeterminacy that results
from the well-known fact that these models are not identified, 
and to make them more applicable in practice, 
we propose two dominance-based normative (or welfare-motivated)
refinements of the set of 
behavioral primitives that are compatible with these models on a given 
dataset. 
The criterion behind both refinements is the respective primitives'
proximity to rational choice: other things equal, 
the more complete either of the two rationales in $\mathbf{RS}$ is, 
and the more alternatives are paid attention to in a given menu, 
the closer the relevant primitive is to rational choice. 

After recalling, in Section 2, the choice models that we focus on, 
in \Cref{subsec:cp-primer} and \Cref{subsec:cdp-primer} we provide brief primers 
of CP and CDP, respectively, that are tailored to the specific optimization and 
model-primitive selection problems that pertain to these models. 
In a nutshell, CP methods---developed by researchers working at
the intersection of optimization theory, computer science, artificial 
intelligence and operations research--- allow us to exploit logical
inference during search. Such inference is crucial as it avoids 
brute-force exploration of an exponentially large space of 
candidate model primitives, which we wish to allow to be either 
\textit{perfectly or approximately compatible} with the choice data
under some model. 
CDP methods, in turn, provide a low-level mechanism for enumerating 
only those perfectly or approximately compatible model primitives  
that are \textit{undominated} according to a dominance relation of interest,
thereby avoiding a separate generate-and-test post-processing step.\footnote{
The specific dominance relations between shortlisting 
and limited-attention model primitives that we introduce and apply 
are dictated by which instance is closer to the model of rational choice, 
other things equal.} Introducing CP and CDP therefore 
enables us to look deeper into the above prominent yet computationally demanding 
models of bounded-rational choice than has been possible under more conventional 
methods thus far. 

In Section 5 we illustrate the usefulness 
of our C(D)P framework (which, we emphasize, 
can be adapted to the particular characteristics of \textit{any} choice-theoretic model 
that may be defined in a general choice environment) with an empirical 
application. There, in addition to assessing the models' explanatory power 
and permissiveness using gold-standard criteria in revealed-preference 
analysis, such as model-based interpretations of the \cite{houtman-maks85} 
proximity index, Bronar's (\citeyear{bronars87}) power test, and Selten's 
(\citeyear{selten91}) measures of predictive success, 
we distill these models' 
welfare-relevant content by enumerating the refined sets 
of ``undominated''
behavioural primitives that are compatible with 
a given dataset under each model.
More specifically, we pin down the full explanatory, predictive and welfare-relevant content
of \textbf{RS}, \textbf{CLA} and \textbf{OC} \textit{vis-\`{a}-vis} \textbf{RC}, 
by applying our C(D)P methods and tool chain on two experimental 
datasets with riskless choices from 4 alternatives/11 menus 
and 5 alternatives/16 menus \citep{manzini23}.\footnote{With the most efficient 
implementation of our CDP models, all computations on these two 
datasets were completed 
within a few hours on a regular laptop computer.}
The majority of subjects' choices in both datasets 
are perfectly compatible with \textbf{RC}. 
Focusing on those that are not, 
we find that the two limited-attention models 
(\textbf{CLA}, followed by \textbf{OC}) explain more subjects' 
choices than \textbf{RS}, but are also considerably more permissive.\footnote{
As usual, permissiveness is quantified via the proportion of simulated 
uniform-random choice datasets that are also compatible with a model.}
Strikingly, our refinement criteria often filter out, respectively, 
thousands and hundreds of thousands of perfectly compatible but 
dominated model primitives in the 4- and 5-alternative datasets. 
In addition, the number of 
undominated such primitives of \textbf{CLA} and \textbf{OC} 
generally exceeds the corresponding number of 
undominated primitives in 
the most structured version of \textbf{RS}, 
which in turn exceeds 
the typically unique compatible preference ordering via which 
\textbf{RC} provides an \textit{approximate} fit for these subjects.
Our analysis, therefore, also uncovers and quantifies an 
interpretational trade-off that researchers of deterministic 
discrete choice are likely to be faced with when 
attempting to understand imperfectly rational behavior: 
perfect fit may come at the cost of poor identification.

From the extensive literature on bounded-rational choice with two-stage models 
since \cite{manzini-mariotti07}, \cite{masatlioglu-nakajima-ozbay12} and 
\cite{lleras-masatlioglu-nakajima-ozbay}, we 
highlight the following contributions as those that are closest in spirit 
to the present paper's focus and questions: 
\cite{declippel-rozen21,inoue-shirai} on testing such models 
when data are limited; \cite{freer-nosratabadi-a,freer-nosratabadi-b,richter20}
on studying non-identification in such models and formulating alternative ones
that help alleviate it; \cite{rubinstein-salant12} on recovering an individual's 
welfare-relevant preferences from imperfectly rational choices generated by 
structured procedures; and \cite{demuynck11,declippel-rozen21} 
on the computational complexity of testing two-stage models with possibly limited
datasets. 
Compared to the most closely related of the studies above, the present one 
is more demanding in its assessment of the underlying models,  
in the sense that it performs such assessment in the models' full 
generality, and aims not only for the kinds of pass/fail checks 
that are afforded by the axiomatic approach, 
but also for: (i) computing the models' \textit{proximity} to a 
dataset; (ii) \textit{enumerating all} possible primitives of the 
model that achieve the closest such proximity; (iii) understanding 
which model fits could not have been achieved under random behavior; 
(iv) comparing models in terms of their predictive success. 

Although motivated by the specific characteristics of these very 
influential models, and delivering their first set of in-depth empirical 
investigations, we view our methodological contribution primarily as 
forward-looking and closely related to 
contemporary studies on over-fitting, under-fitting 
and complexity trade-offs in model selection, 
e.g. \cite{liang-JLE,fudenberg-lian-gao21,fudenberg-gao-you,frechette-vespa-yuksel26,
andrews-fudenberg-lei-liang-wu,fudenberg-kleinberg-liang-muillainathan}.
Unlike the focus of these studies on parametric models of risk preference
or belief updating, ours is a contribution to this literature on the domain 
of non-parametric and deterministic discrete choice models.
In particular, we hope that constraint (dominance) programming 
will be a useful new framework in behavioral revealed preference analysis 
that will facilitate the development of a 
new generation of models in this domain that will also 
be informed by the---now measurable---trade-offs in their degrees of 
perfect or approximate explanatory power; indeterminacy; and predictive success.  

\section{Rational Choice, Shortlisting and Limited Attention}

\subsection{Definitions}

The finite set that contains all choice alternatives is denoted by $X$. 
To avoid repetition, in the definitions pertaining to 
properties of a binary relation $\succ$ on $X$ 
the relevant requirement is assumed to hold for all 
items in $X$. 
Specifically, such a relation $\succ$ is said to be 
\textit{irreflexive} if $x\nsucc x$; 
\textit{asymmetric} if $x\succ y$ $\Rightarrow$ $y\nsucc x$; 
\textit{transitive} if $x\succ y \succ z$ $\Rightarrow$ $x\succ z$; and  
\textit{total} if $x\neq y$ $\Rightarrow$ $x\succ y$ or $y\succ x$.
An irreflexive relation $\succ$ is \textit{acyclic} 
if $x_1\succ x_2 \succ \cdots \succ x_k$ $\Rightarrow$ $x_k\nsucc x_1$. 
A relation $\succ$ is a \textit{strict partial order} if it 
is asymmetric and transitive. 
A total strict partial order is a \textit{strict linear order}.
Strict partial orders are acyclic relations, and the latter are asymmetric.
The converse statements are false. 

	A single-choice dataset on $X$ (henceforth simply \textit{dataset}) 
	is a tuple $\mathcal{D}=\{(A_i,C(A_i)\}_{i=1}^k$ where, for all 
	$i\leq k$, $C(A_i)\subseteq A_i\subseteq X$ is a singleton
	and $A_i\neq A_j$ whenever $i\neq j$. 
	A dataset $\mathcal{D}$ is explained by  
	\textnormal{rational choice with strict preferences (\textbf{RC})} 
    if there is a strict linear order $\succ$ on $X$ such that, 
	for all $i\leq k$,
		\begin{eqnarray}
			\label{RC}	C(A_i) & = & M_{\succ}(A_i),
		\end{eqnarray}
\noindent	where, for any $A\subseteq X$,
	\begin{eqnarray*}
		M_{\succ}(A) & := 
		& \{x\in A: y\not\succ x \text{ for all } y\in A\}.\
	\end{eqnarray*}
\noindent 
The set $M_\succ(A)$ is the set of $\succ$-\textit{undominated}, 
or $\succ$-\textit{maximal}, alternatives in $A$. 
When $\succ$ is a strict linear order, $M_\succ(A)$ consists of 
a single item, which is also the \textit{dominant} or \textit{greatest} 
element of $\succ$ in that set. 
If $\succ$ is acyclic and not merely assymmetric, then $M_\succ(A)$ is non-empty.
A dataset $\mathcal{D}$ is explained by the 
	\textnormal{``rational shortlist method'' (\textbf{RS})} \citep{manzini-mariotti07}
    if there exist asymmetric relations 
	$\succ_1$ and $\succ_2$  
	on $X$ such that, for all $i\leq k$,
		\begin{eqnarray}
		\label{SRC-2}	C(A_i) & = & M_{\succ_2}\big(M_{\succ_1}(A_i)\bigr)
		\end{eqnarray}
A dataset $\mathcal{D}$ is explained by 
	\textnormal{``choice with limited attention'' (\textbf{CLA})}
	\citep{masatlioglu-nakajima-ozbay12}
	if there is a pair $(\Gamma,\succ)$ 
	where $\succ$ is a strict linear order on $X$ 
	and $\Gamma:2^X\setminus\{\emptyset\}\rightarrow 2^X\setminus\{\emptyset\}$ 
	an \textnormal{``attention filter''} such that, for all $i\leq k$,
	\begin{subequations}
		\begin{eqnarray}
			\label{AF-1a} \Gamma(A_i) 
			&  \subseteq   & A_i\\
			\label{AF-2} x\not\in \Gamma(A_i) 
			& \Longrightarrow & \Gamma(A_i)=\Gamma(A_i\setminus\{x\})\\
			\label{AF-5} C(A_i) & = & 
			M_\succ\big(\Gamma(A_i)\bigr)
		\end{eqnarray}
	\end{subequations}	
Finally, a dataset $\mathcal{D}$ is explained by
\textnormal{``overwhelming choice'' (\textbf{OC})}
	if there is a pair $(\Gamma,\succ)$ where $\succ$ is a strict linear 
	order on $X$ and 
	$\Gamma:2^X\setminus\{\emptyset\}\rightarrow 2^X\setminus\{\emptyset\}$ 
	a \textnormal{``competition filter''} such that, for all $i\leq k$, 
	\eqref{AF-1a} and \eqref{AF-5} hold while \eqref{AF-2} is 
	replaced by
	\begin{eqnarray}
		\label{OC-2} x\in A_i \subset A_j \text{ and } x\in \Gamma(A_j)
		& \Longrightarrow & x\in \Gamma(A_i)
	\end{eqnarray}
\noindent 
Replacing \eqref{AF-2} with \eqref{OC-2}, and vice versa, makes 
\textbf{CLA} and \textbf{OC} logically distinct. 
The former condition is an inattention-irrelevance one:
if an option is not considered at a menu, then whether or not it is 
feasible there does not affect the set of options that are 
actually paid attention to. This may be an appealing property
in some situations but does not allow for thinking about cases 
where the only reason why some alternatives are not considered 
is the menu's off-puttingly big size. In other words, these options 
may well be paid attention to if they remain feasible in a subset 
of the original ``big'' menu that is now sufficiently small 
so as to be cognitively manageable by the agent.
This is allowed by \eqref{OC-2}, though not by \eqref{AF-2}. 
	
\paragraph{Example (Part 1).}

Consider the following dataset from an underlying set $X$ consisting of 
$x$, $y$ and $z$:
\begin{eqnarray}
\nonumber 
\label{eq:example-data}
\mathcal{D} & = & \Big\{\big(A_i,C(A_i)\bigr)\Bigr\}_{i=1}^4\\
& = & 
\Big\{
\big(\{x,y\},x\bigr),  
\big(\{y,z\},y\bigr),  
\big(\{x,z\},z\bigr),  
\big(\{x,y,z\},x\bigr) 
\Bigr\}
\end{eqnarray}
Since $\mathcal{D}$ features a binary choice cycle, it is incompatible
with \textbf{RC}. 
It is, however, explainable by \textbf{RS}, \textbf{CLA} and \textbf{OC}, 
as follows (we omit the simple verification details):

\noindent 
\textbf{RS}: $(\succ_1, \succ_2)$ defined by 
$y\succ_1 z$ and the strict linear order 
$z\succ_2 x \succ_2 y$.

\noindent \textbf{CLA}: 
$(\Gamma^{CLA},\succ^{CLA})$ defined by 
$x\succ^{CLA} y\succ^{CLA} z$ and the attention filter $\Gamma^{CLA}$ 
such that 
$(A_1,A_2,A_3,A_4)\xrightarrow[]{\Gamma^{CLA}}(\{x,y\}, \{y,z\}, \{z\}, \{x,y\})$.

\noindent \textbf{OC}: $(\Gamma^{OC},\succ^{OC})$ defined by 
$y\succ^{OC} z\succ^{OC} x$ and the competition filter $\Gamma^{OC}$ 
such that  $(A_1,A_2,A_3,A_4)\xrightarrow[]{\Gamma^{OC}} (\{x\}$, $\{y,z\}$, $\{x,z\}$, $\{x\})$.

Yet these three instances of \textbf{RS}, \textbf{CLA} and \textbf{OC} 
are not the only ones 
that explain these choices. 
Applying our CP method we 
enumerate in Appendix A all 
12 distinct pairs $(\succ_1,\succ_2)$ of asymmetric relations that are compatible 
with them under \textbf{RS}, as well as all 
11 and 6 pairs $(\succ,\Gamma)$ of strict linear orders
and attention/competition filters that are so 
under \textbf{CLA} and \textbf{OC}, respectively. 
\hfill $\blacklozenge$

\subsection{Decomposition of Shortlisting into Three Nested 
	Classes\label{subsec:decomposition}}

Although we study three logically distinct models of bounded-rational choice, 
it will prove more insightful in the sequel to decompose \textbf{RS} 
into three nested classes, depending on the structure imposed 
on $\succ_1$ and $\succ_2$. 
The most general one clearly corresponds to both  
$\succ_i$ being asymmetric and possibly cyclic, i.e.   
the one axiomatized in \cite{manzini-mariotti07}.
We will refer to this as \textbf{RS}$_{asym}$. 
But special cases of natural interest here are also those where both 
$\succ_i$ are either acyclic (\textbf{RS}$_{acyc}$) or 
strict partial orders (\textbf{RS}$_{order}$). 
In fact, because of their relative tractability, the latter have 
received more attention in the choice-theoretic and 
revealed-preference literatures 
\citep{rubinstein_salant_hndbk,au-kawai11,dutta-horan15,inoue-shirai,declippel-rozen21}. 

Clearly, we have
\begin{eqnarray}
	\label{eq:nests}
	\textbf{RS}_{asym}\supset\textbf{RS}_{acyc}
\supset\textbf{RS}_{order}
\end{eqnarray}
because \textbf{RS}$_{asym}$ contains 
pairs $(\succ_1,\succ_2)$ where 
at least one of the two asymmetric relations may be cyclic 
and \textbf{RS}$_{acyc}$ contains pairs where at least one of 
the two acyclic relations may not be transitive. 
Using this finer taxonomy, we can now clarify that  
out of this model's 12 primitives that are perfectly compatible with 
the above example choices (Appendix A), one is in the \textbf{RS}$_{order}$ class; 
nine in the \textbf{RS}$_{acyc}\setminus$\textbf{RS}$_{order}$ class 
(i.e. both relations in each primitive are acyclic and at least one
is not transitive);  
and two in the \textbf{RS}$_{asym}\setminus$\textbf{RS}$_{acyc}$ class
(i.e. both relations in each primitive are asymmetric and at least one
is cyclic).

\section{Computing Models' Proximity to the Data}

{

As we highlighted earlier, we are not merely interested 
in conducting a ``pass-or-fail'' 
test for each model, but also in assessing ``how close'' 
that model is to explaining a dataset 
where its fit is possibly imperfect. 
A standard such metric in revealed preference analysis,  
applicable both on discrete feasible sets and continuous/divisible ones that are determined
by a budget constraint, is the so-called \cite{houtman-maks85} (HM) score
\citep{choi-fisman-gale-kariv07,smeulders-spieksma-cherchye-derock,
caplin-dean-martin11,heufer-hjertstrand15,apesteguia-ballester15,
dean-martin16,CCGT22,gerasimou26,demetry-hjertstrand23,demuynck-rehbeck23}.
In its original formulation, this is defined as the smallest number of choices that 
need to be changed in (or, equivalently, dropped from) a subject's dataset in order for 
them to be compatible with the model of rational choice under some preference ordering. 
However, this  principle can be applied to any model, choice heuristic 
or axiom.\footnote{``Prest'' \citep{gerasimou-tejiscak-prest} is a free and 
open-source desktop tool that uses brute force optimization to compute this 
measure for several bounded-rational choice models of preference maximization 
that are defined only in terms of a---possibly incomplete---preference ordering.
The measure itself is intuitive and natural enough that other papers 
in behavioral revealed preference---for example, \cite{ellis-freeman24}---report 
it without reference to \cite{houtman-maks85}.}
In such cases, a model's HM score---more descriptively, 
\textit{distance score}---on a dataset similarly corresponds to the number 
of changes that are necessary for that model to fit those data perfectly 
under some of its permissible instances. 

Formally, for each $\mathcal{D}$, the score of a model belonging to one 
of the $\mathbf{RS}_{class}$, $\mathbf{CLA}$ or $\mathbf{OC}$ classes is 
defined, respectively, by 
\begin{eqnarray}
	\label{score1}
	Score\big(\mathbf{RS}_{class}(\mathcal{D})\bigr)&=&
	\min\limits_{(\succ_1,\succ_2)\in\mathbf{RS}_{class}}
	\{|\{(A_i,C(A_i)\in\mathcal{D}\}|:C(A_i)\neq M_{\succ_2}\big(M_{\succ_1}(A_i)\bigr)\}
	\qquad
	\\
	\nonumber &&\\
	\label{score2}
	Score\big(\mathbf{CLA}/\mathbf{OC}(\mathcal{D})\bigr)&=&
	\min\limits_{(\Gamma,\succ)\in\mathbf{CLA}/\mathbf{OC}}
	\{|\{(A_i,C(A_i)\in\mathcal{D}\}|:C(A_i)\neq M_{\succ}\big(\Gamma(A_i)\bigr)\}
\end{eqnarray}
An intuitive interpretation of a model's distance score for a given dataset
is that it counts how many of the relevant decision maker's
choices might be considered ``mistakes'' from the 
perspective of that model, with a zero score reflecting a perfect model fit. 

Computing the three models' distance scores defined in 
\eqref{score1} and \eqref{score2}---let alone enumerating the full 
set of solutions in the respective $\arg\min$ sets, which we also 
set out to do--- 
is computationally hard, owing to the 
fact that their primitives grow exponentially in the size of the underlying set
of alternatives.\footnote{Computing the HM score for \textbf{RC} is also hard.
See \cite{smeulders-spieksma-cherchye-derock} for an analysis in the neoclassical
demand-theoretic environment.} 
To give a sense of this we note that the numbers of distinct 
strict linear orders, strict partial orders, acyclic and asymmetric relations 
grow from 24, 219, 543 and 2,680 when $|X|=4$ to 120, 4,231, 29,281 and 109,824 
when $|X|=5$.\footnote{Recall that the number of strict linear orders on $X$ 
is $|X|!$. For the latter three types of binary relations see the 
\href{https://oeis.org/A001035}{A001035}, 
\href{https://oeis.org/A003024}{A003024} and \href{https://oeis.org/A123553}{A123553} entries
of the Online Encyclopedia of Integer Sequences in \href{https://oeis.org}{https://oeis.org}.}
This makes brute-force search for the models' distance scores, 
and recovery of the set of model instances that are compatible with 
these scores, an impractical approach. When $|X|=4$, 
for example, using this method to test the most general version of the shortlisting model 
against a single subject's dataset would involve searching $2,680^2=7,182,400$ pairs 
of asymmetric binary relations $(\succ_1,\succ_2)$\footnote{
That said, 
we note that \citet[Theorem 3]{demuynck11} constructs an algorithm to establish 
that a pass-or-fail test of \textbf{RS} is possible in polynomial time on $|X|$.} 
and 630,118,440 pairs 
$(\succ,\Gamma)$ of strict linear orders and attention/competition filters\footnote{Indeed, 
there are 11 non-singleton menus when $|X|=4$: 6 binary; 4 ternary; 
1 quaternary. For binary menu $\{x,y\}$ the possible values of $\Gamma(\{x,y\})$ are 3: 
$\{x\}$, $\{y\}$ and $\{x,y\}$. Similarly, for ternary menu $\{x,y,z\}$ and quaternary menu
$\{w,x,y,z\}$, these are 7 and 15, respectively. Thus, given that there are 24 strict linear
orders on a set with 4 elements, it follows that there are 
$24\times 3^6\times 7^4 \times 15^1=630,118,440$ 
pairs $(\succ,\Gamma)$.} toward
testing either limited-attention model. 

\subsection{A Primer in Constraint Programming\label{subsec:cp-primer}}

The computational tasks in this paper are not standard problems where a real- or integer-valued 
objective function is to be minimized subject to some constraints.
Had this been the case, we could have resorted to using tools for 
solving mixed integer linear programs (MILP).\footnote{For a recent 
demonstration of integer-programming methods in revealed preference
analysis from budget-constrained, consumer-theoretic choices we 
refer the reader to \cite{demuynck-rehbeck23}.}
Rather, we repeatedly need to: (i) decide whether there exist primitives---i.e.
$(\succ_1,\succ_2)$ or $(\succ,\Gamma)$---consistent with a dataset after
dropping some observations, as part of the process of computing the scores in 
\eqref{score1} and \eqref{score2}; 
and (ii) \textit{enumerate} the set of
primitives that achieve a given score for a model, 
possibly after applying a dominance-based 
refinement (we return to the latter point after introducing 
Definitions \ref{dfn:dominance-rs} and \ref{dfn:dominance-la}, below). 
This is precisely the type of setting for which constraint-based approaches 
are designed \citep{rossi-vanbeek-walsh06,apt03,dechter03}.

We start by stressing that CP separates modelling from solving.
Specifically, in CP one first writes a \textit{model} of a problem
in a bespoke, human-friendly programming language. 
This model comprises
a finite collection of decision \textit{variables} 
(the unknown objects to be recovered); their \textit{domains} 
(the admissible values of those objects); and \textit{constraints} 
(the logical conditions they must satisfy) \citep{apt03}.
A CP \textit{solver} then searches for assignments to the decision variables that
satisfy all constraints. Importantly, the model is conceptually distinct from the
algorithm used to solve it: the same high-level model can be compiled to different
low-level solver formalisms, and solved by 
different backends/solving engines.
In our implementation we first 
write models in the \textsc{Essence} language 
\citep{frisch2008essence} (full details are available in Appendix C). 
We then use the tools \textsc{Conjure} and \textsc{Savile Row} 
to translate these models into solver-optimal input 
\citep{akgun2022conjure,SavileRow}. 
Finally, this input is passed to the solver for execution
and output production.
We report our results with the \textsc{Minion} solver
\citep{MINION}, which is explicitly designed for combinatorial-optimization
problems such as the ones we tackle in this paper. We stress, however, that our }
modelling pipeline is not tied to a single solver: changing
the backend is (conceptually and, in practice, often operationally) 
a compilation choice rather than a re-modelling exercise.
In other words, the same human-friendly and choice-model specific 
code---i.e. in our case, those written in \textsc{Essence} 
and shown in A.C.---can
be sent to different backend solvers at no cost to the researcher.

A useful way to view CP solving is as an interleaving 
of two components: \textit{inference} (also called constraint propagation) 
and \textit{search} \citep{bessiere06,vanbeek06}.
This \textit{``embeds any reasoning which consists in 
explicitly forbidding values or combinations of values for some variables 
of a problem because a given subset of its constraints cannot be
satisfied otherwise''} \citep[p.26]{bessiere06}.
For example (taken from the same source), if a problem requires that 
integer variables $x_1,x_2$ must lie in the set $\{1,\ldots,10\}$ 
and a constraint asks that $|x_1-x_2|>5$, then propagating this constraint
leads to the inference that $x_1,x_2\not\in \{5,6\}$. 
More generally, inference reasons from the constraints to rule out variable 
values that cannot participate in any feasible solution, while 
search proposes choices (partial assignments) when inference alone 
does not determine a unique outcome.

In practice, solvers
implement this inference via \textit{filtering algorithms}, including specialised
procedures for widely used constraint families (often called \textit{global constraints})
\citep{regin04}.
The solver alternates
between these: it makes a tentative assignment decision; propagates implications;
detects contradictions early; and \textit{backtracks} when necessary. 
This is the key mechanism by which CP avoids naive brute-force enumeration 
of all possibilities \citep{bessiere06,vanbeek06,mackworth77}.
In the language of the CP literature, propagation \textit{prunes} domains 
while backtracking organises the exploration of the remaining space.

\paragraph{Example (Part 2).}

To illustrate how inference/constraint propagation works in our setting, 
consider the example data introduced earlier. 
Under the \textbf{RS} model, constraint propagation with respect 
to the first rationale, $\succ_1$, of some candidate instance $(\succ_1,\succ_2)$
of the model that may explain the given choices amounts to reasoning that
$$
\begin{array}{lll}
	C(\{x,y\})=x & \Rightarrow & y \nsucc_1 x\\
	C(\{y,z\})=y & \Rightarrow & z \nsucc_1 y\\
	C(\{x,z\})=z & \Rightarrow & x \nsucc_1 z\\
	C(\{x,y,z\})=x & \Rightarrow & y \nsucc_1 x\; \& \; z \nsucc_1 x 
\end{array}\quad 
\Longrightarrow \quad
\begin{array}{lll}
	x\nsucc_1 z \nsucc_1 x && (\text{C}1)\\
	y\nsucc_1 x && (\text{C}2) \\
	z\nsucc_1 y && (\text{C}3) 
\end{array}
$$
From the exclusions imposed by constraints (C1)--(C3), then,  
it is inferred that the possible relations $\succ_1$ 
for any candidate instance $(\succ_1,\succ_2)$ are:
\begin{enumerate}
	\item $y\succ_1 z$ \quad 
	{\hfill (compatible with \textbf{RS}$_{asym}$, 
		\textbf{RS}$_{acyc}$, \textbf{RS}$_{order}$)}
	\item $x\succ_1 y$ \quad 
	{\hfill (compatible with \textbf{RS}$_{asym}$, 
		\textbf{RS}$_{acyc}$, \textbf{RS}$_{order}$)}
	\item $x\succ_1 y$, $y\succ_1 z$ \quad 
	{\hfill (compatible with \textbf{RS}$_{asym}$, \textbf{RS}$_{acyc}$)}
\end{enumerate}
Constraint propagation/inference therefore  
leads to a meaningful computational gain by reducing 
the search for a compatible instance $(\succ_1,\succ_2)$
and only retaining either 2 or 3 of the 128 possible relations 
$\succ_1$ in this search.
\hfill $\blacklozenge$\\

Many CP applications aim to find a single feasible solution or a single optimum.
As already highlighted, our setting additionally requires 
\textit{enumeration}: counting and listing all
primitives that are compatible with a given model distance score, and then subjecting them to
further refinement criteria. CP supports enumeration naturally: once a solution is
found, the solver can be instructed to continue searching for additional solutions.
Operationally, this is typically implemented by repeatedly solving while adding a
\textit{blocking constraint} that excludes the solutions already found or a
region of the solution space implied by those solutions. In our application this
supports the tasks of interest: ``enumerate all $(\succ_1,\succ_2)$ or 
$(\succ,\Gamma)$ pairs that achieve the
minimum score''.

Finally, a practical advantage of the particular set of free and open source 
CP tools that we deploy here, 
comprising \textsc{Essence} together with
\textsc{Conjure} and \textsc{Savile Row}, is that it lets us keep the economic
decision-making structure of the problem explicit 
(relations, sets, and menu-indexed objects) while leaving the
low-level encoding choices to the modelling tool chain. 
This is valuable in a study 
like ours where it is useful to be able to change the backend solver without 
rewriting the economic model, and where the same conceptual constraints 
are reused across: (i) model distance scoring; (ii) exact compatibility;
(iii) dominance-refined enumeration.

\section{Eliciting the Models' Welfare-Relevant Content}

We now use these computational tools to address the central
economic question of welfare inference in the presence of 
imperfectly rational choice data. 
To this, we note that it is well-known---and already alluded to in the first part
of our running example---that 
the \textbf{RS}, \textbf{CLA} and \textbf{OC} 
models are not identified: even when choices from the full domain of menus are available to the analyst, 
there is generally more than one instance of each model that could explain the data 
if \textit{some} instance of the model does so. 
Aware of the possibility that the same data can be explained by multiple models, 
some authors have suggested the use of non-choice data in an attempt to 
understand which one(s) might be the most relevant for the problem of interest.
But what if such additional non-choice data are not available? 
And, more pertinent to our study, how should an analyst 
decide which one(s) among the multiple instances within the same model is/are
the most appropriate for welfare analysis, irrespective of whether such 
additional non-choice data are available or not? We turn to this question next.

\subsection{Proximity to Rational Choice as A Refinement Criterion}

Multiplicity of a model's instances that provide an equally good explanation 
of a decision maker's choices can be either genuine, in the sense that the models'
primitives across these instances are in conflict with each other, 
or mechanical, in the sense that those primitives differ in lesser, 
non-conflicting ways. 
To distinguish the former from the latter type of indeterminacy, we make use 
of the fact that all three models of bounded-rational choice in our focus 
include \textbf{RC} as a special case, 
and introduce a \textit{dominance filtering} criterion that favors those 
instances of each model that are ``closer'' to \textbf{RC}.
In the case of \textbf{RS}, this means favoring instances that contain a ``more
complete'' relation $\succ_1$ or $\succ_2$, \textit{ceteris paribus}.
In the case of \textbf{CLA} and \textbf{OC}, it means favoring instances featuring
larger consideration sets, \textit{ceteris paribus}. 

\begin{dfn}
	\label{dfn:dominance-rs}
	If $(\succ_1^a,\succ_2^a)$ and $(\succ_1^b,\succ_2^b)$ are two distinct 
	instances of a shortlisting model \textnormal{\textbf{RS}}$_{class}$, 
	$class\in\{order,acyc,asym\}$,
	that explain a dataset $\mathcal{D}$, then the former is said to 
	\textnormal{dominate} the latter \textnormal{in \textbf{RS}$_{class}$} if
	
	\vspace{-30pt}
	 
	\begin{subequations}
		\begin{eqnarray}
		\label{dom-RSM-1}	
		\succ_t^b & \subseteq & \succ_t^a \qquad \text{ for all } t\in\{1,2\}\\
		\label{dom-RSM-2}	
		\succ_t^b & \subsetneq & \succ_t^a \qquad \text{ for some } t\in\{1,2\}
		\end{eqnarray}
	\end{subequations}
	
	\vspace{-18pt}
	
	\noindent
	An instance in \textnormal{\textbf{RS}}$_{class}$ is undominated 
	if it is not dominated in \textnormal{\textbf{RS}}$_{class}$.
\end{dfn}

This notion is applicable as a filtering criterion 
within each of the three logically nested \textbf{RS} classes
[cf. \eqref{eq:nests}], 
which, as was noted in \Cref{subsec:decomposition}, are defined 
by the consistency structure 
of the least consistent rationale in the relevant pair/instance of the model. 
Finding the undominated compatible pairs/instances within a given \textbf{RS} 
class, however, does not immediately suggest a way of 
comparing those instances \textit{across} classes. 
For example, the pairs $(\succ_1^a,\succ_2^a)$ and $(\succ_1^b,\succ_2^b)$
defined by 
$$
\begin{array}{cccc}
    & \succ_1^\star          && \succ_2^\star\\
\star=a:  & (x,y)       && (z,y),(y,x),(z,x)\\  
\star=b:  & (x,y),(y,z) && (z,y),(y,x),(z,x)
\end{array}
$$
correspond to shortlisting instances in the \textbf{RS}$_{order}$
and \textbf{RS}$_{acyc}$ classes, respectively, that
feature an identical second rationale. 
If they were treated as belonging to the same class, 
the latter instance would clearly dominate the former. 
A question now arises: Should an acyclic but intransitive shortlisting 
relation be considered closer to being rational than a transitive such 
relation that contains strictly fewer ranked comparisons?
More generally, how should one think about trade-offs involving 
the hierarchical consistency of rationales---whereby 
$transitive \gg acyclic \gg asymmetric$ in this order---and their plurality?

We imagine both approaches being potentially appropriate 
in some contexts. 
Prioritizing consistency may be relatively uncontroversial 
if one of the two relations in the debate is cyclic,
not least because cyclicality can be detrimental to any 
attempt to draw actionable welfare-relevant conclusions. 
This is less so, however, when one 
relation is transitive while the other---although 
merely acyclic---contains more pairs of comparable alternatives, 
as is the case in the above example. 
An obvious argument in favor of plurality here is that the 
information contained in the latter relation is not 
self-contradictory and, being also richer than its opponent, 
may provide more targeted welfare-relevant guidance. 
On the other hand, if it is postulated that $x$ is somehow
better than $y$, and $y$ in turn is considered to be better than $z$, 
yet $x$ and $z$ cannot be compared, a cautious observer 
might be concerned about the validity (or lack thereof) of one 
or both of the two antecedent comparisons. For such an observer
consistency might still be prioritized over plurality. 

\begin{dfn}
	\label{dfn:dominance-la}
	If $(\succ^a,\Gamma^a)$ and $(\succ^b,\Gamma^b)$ are two distinct 
	instances of a limited-attention model that explain a dataset $\mathcal{D}$,
	then the former is said to \textnormal{dominate} the latter if 
	\begin{subequations}
		\begin{eqnarray}
			\label{dom-LA-1}	
			\succ^b & = & \succ^a\\
			\label{dom-LA-2}	
			\Gamma^b(A_i) & \subseteq & \Gamma^a(A_i) \qquad \text{ for all } i\leq k\\
			\label{dom-LA-3}	
			\Gamma^b(A_i) & \subsetneq & \Gamma^a(A_i) \qquad \text{ for some } i\leq k
		\end{eqnarray}
	\end{subequations}
\end{dfn}
That is, a data-compatible limited-attention model primitive dominates another 
such primitive if they both feature the same underlying preference relation but
the former portrays the decision maker as always paying attention to at least as
many choice alternatives as the latter, and in at least one case 
as paying attention to strictly more alternatives.
One can imagine an additional filtering process being applied among the primitives
that survive this refinement, where those featuring 
attention/competition filter mappings that are finer in the sense 
of \eqref{dom-LA-2}--\eqref{dom-LA-3} are favored even when the underlying 
preference relations differ. While easily implementable in our setting, 
such an additional filtering process is not without loss of potentially 
welfare-relevant information, unlike the one put forward in Definition 
\ref{dfn:dominance-la}.

\paragraph{Example (Part 3).}

As noted earlier, Appendix A shows the  
12, 11 and 6 distinct
primitives of the \textbf{RS}, \textbf{CLA} and \textbf{OC} 
models that are compatible with the cyclic choices in 
our running example.
Appendix B explains why: 3 of the 9 compatible primitives 
$(\succ_1,\succ_2)$ in the 
$\mathbf{RS}_{acyc}\setminus\mathbf{RS}_{order}$ class 
are undominated; 
1 of the 2 in $\mathbf{RS}_{asym}\setminus\mathbf{RS}_{acyc}$;
3 of the 11 \textbf{CLA} primitives $(\Gamma,\succ)$;
and 3 of the 6 \textbf{OC} primitives.
\hfill $\blacklozenge$\\

Intuitively appealing as they may be for the empirical revealed preference 
analysis of bounded-rational choices, it is not obvious how the proposed 
dominance-based refinements 
of the \textbf{RS} and \textbf{CLA}, \textbf{OC} compatible model 
primitives 
can be operationalized so that they are applied systematically in practice.
This is our motivation for introducing and tackling this problem with CDP.

\subsection{A Primer in Constraint \textit{Dominance} 
Programming\label{subsec:cdp-primer}}	

Standard constraints restrict what a \textit{single} 
candidate primitive may look like
(e.g.\ that $\succ$ is a strict linear order, or that $\Gamma$ satisfies the 
attention-filter conditions). These are constraints \textit{on a solution}. 
However, the dominance relations in Definitions \ref{dfn:dominance-rs} and
\ref{dfn:dominance-la} are not properties of a single instance in isolation: 
they are relations \textit{between} instances. 
Such requirements are naturally expressed as constraints \textit{among solutions}.
A direct way to compute undominated instances is therefore a two-step
generate-and-test procedure: 1. Enumerate \textit{all} feasible solutions (all compatible instances);
2. Post-process the resulting list to filter out those that are dominated.
This works, but has a clear computational weakness: it requires a full
enumeration even when a large fraction of the enumerated instances are dominated and
will ultimately be discarded. 

Is there a better alternative?
Indeed, CDP avoids such exhaustive generate-and-test.
Instead, it is a principled way to integrate dominance
filtering into the enumeration process itself 
\citep{guns-stuckey-tack18,chu-stuckey15}. The modeller specifies a dominance
relation over solutions, such as those in Definitions \ref{dfn:dominance-rs} and 
\ref{dfn:dominance-la}, and the CDP mechanism ensures that once a solution is found,
the subsequent search is constrained so as to exclude solutions dominated by it.
Concretely, after discovering a solution $s$, the system posts a \textit{dominance
blocking constraint} that rules out any future solution $s'$ such that $s$ dominates
$s'$. Search then continues in the remaining space, which (by construction) contains
no solution dominated by an already accepted one. The net effect is that the solver
spends its effort on discovering new \textit{undominated} solutions, rather than on
discovering large numbers of dominated ones and discarding them afterwards.

CDP nevertheless does not remove the need for post-processing completely. The reason
is that, during enumeration, a solution that is \textit{non-dominated so far} may later
turn out to be dominated by a solution found subsequently. In such cases, the earlier
solution will have been recorded even though it is not undominated in the final set.
A lightweight post-processing pass can therefore still be required to remove any
solutions that are dominated \textit{eventually} by later discoveries. CDP should thus be
viewed as a way to avoid \textit{exhaustive} generate-and-test (by steering search away
from dominated regions as it progresses), rather than as a guarantee that every
intermediate output is globally undominated.
For the problem at hand, 
CDP can substantially reduce the burden of enumerating
and filtering large families of dominated \textbf{RS}, 
\textbf{CLA} and \textbf{OC} instances, leaving only minimal
post-processing to ensure that the retained set is undominated in the global sense.

\section{Illustration with Choices from Two Experiments}

\subsection{Data}

We use data from two online experiments conducted by \cite{manzini23}.
Participants were recruited on the ``Ravelry'' social network which, according to
Wikipedia (accessed on 26th June 2026) \textit{``functions 
as an organizational tool for a variety of fiber arts, 
including knitting, crocheting, spinning and weaving. 
Members share projects, ideas, and their collection of yarn, 
fiber and tools via various components of the site.''} 
In the first experiment the grand choice set $X$ comprised 4 knitted dresses. 
Subjects (primarily female senior citizens) were asked to 
make a single choice from all 11 menus with 2, 3 or 4 
items that are derivable from that set. 
A well-defined, single-valued choice function was therefore elicited for 
every subject there.
In the second experiment, $X$ comprised 5 knitted ponchos.
Here, subjects made single choices from 16 of the 26 non-trivial 
menus (61.5\%) that are derivable from $X$; specifically, those that 
comprised the grand set itself and those with 2 or 4 items. A partially
defined choice function on the same sub-domain was therefore elicited for every subject.
All subjects were rewarded with a knitting pattern, and a randomly selected subset 
of them also received a yarn in their chosen color to knit with it. 
Out of the 277 and 336 subjects with fully recorded choice responses in these
two experiments, respectively, 195 (70.4\%) and 169 (50.3\%)
were explained perfectly by rational choice with strict preferences. 
Our primary focus will therefore be on the remaining 82 
and 167 subjects in these ``Four Dresses'' (4D) and ``Five Ponchos''
(5P) data.

\subsection{Explanatory Power and Predictive Success}

Table \ref{tab:model-scores} presents the results of our non-parametric 
model-score analysis in the sense of \eqref{score1}-\eqref{score2}
of the non-\textbf{RC}-compliant subjects in the 4D 
and 5P data (panels (a) and (b), respectively).
In the former data, 77 of the 82 subjects (94\%) are perfectly 
explainable by at least one of the \textbf{RS}, \textbf{CLA} 
and \textbf{OC} models. Among them, 75 and 51 are 
perfectly compatible with \textbf{CLA} and \textbf{OC}, respectively,
with all remaining ones being approximately compatible, with a score of 1. 
Furthermore, 4 and 53 subjects, respectively, are perfectly and 
approximately compatible---with a score of 1---with all 
three classes of the \textbf{RS} model.
Similar results are obtained in the larger Five-Ponchos sample. 
Notably, \textbf{CLA} now explains \textit{all} subjects
perfectly, while the proportion of those explained 
by \textbf{RS} is doubled compared to 4D
(for \textbf{OC} they are the same). 
Additionally, these data allow for differentiating 
between the explanatory power of the more general 
\textbf{RS}$_{acyc}$ and \textbf{RS}$_{asym}$ classes 
relative to the transitive \textbf{RS}$_{order}$ class, with the 
former accounting perfectly for one additional subject's behavior.

\begin{table}[!htbp]
\small
	\caption{\centering
	Proportions of subjects who are incompatible with Rational Choice 
	and are perfectly or imperfectly 
	explained by the Shortlisting and Limited-Attention models.}
	\makebox[\linewidth][c]{
	\setlength{\tabcolsep}{5pt} 
	\renewcommand{\arraystretch}{1.2} 
	\begin{tabular}{|r|c|c|c|c|c|}
		\hline
		\textit{Perfect \&}
		&\textbf{RS\textit{order}}
		&\textbf{RS\textit{acyc}}
		&\textbf{RS\textit{asym}}
   		&\textbf{OC}
		&\textbf{CLA}\\
		\cline{2-6}
		\textit{approximate fits}
		& \multicolumn{5}{c|}{Four Dresses ($N=82$)}\\
		\hline
		Score=0
		&4.8\%
		&4.8\%
		&4.8\%
		&62.2\%
        &91.5\%\\
		\hline
		Score=1 
		&64.6\%
		&64.6\%
		&64.6\%
		&37.8\%
		&8.5\% \\
		\hline
		Score=2
		&20.7\%
		&21.9\%
		&21.9\%
		&--
		&-- \\
		\hline
		Score=3
		&9.9\%
		&8.7\%
		&8.7\%
		&--
		&-- \\
		\hline
		& \multicolumn{5}{c|}{Five Ponchos ($N=167$)}\\
		\hline
		Score=0
		&8.9\%
		&9.6\%
		&9.6\%
		&62.3\%
		&100\% \\
		\hline
		Score=1 
		&55.7\%
		&55.7\%
		&55.7\%
		&37.7\%
		& -- \\
		\hline
		Score=2
		&23.5\%
		&25.1\%
		&25.1\%
		& --
		& -- \\
		\hline
		Score=3
		&8.9\%
		&8.4\%
		&8.4\%
		& --
		& -- \\
		\hline
		Score$\geq$4
		&3.0\%
		&1.2\%
		&1.2\%
		& --
		& -- \\
		\hline
	\end{tabular}
	}
	\label{tab:model-scores}
\end{table}

Next, following the ideas put forward in 
\cite{bronars87,selten91} 
and later refined in  \cite{beatty-crawford11}, 
\cite{cherchye-demuynck-derock-lanier25} and \cite{fudenberg-lian-gao21}, 
among many others, we assess the predictive 
ability of the three models in the 4D and 5P data 
by juxtaposing how well they explain 
human vs. synthetic subjects who choose uniform-randomly
in the respective collections of 11 and 16 menus. 
In particular, for a model $M$ in
$\{\mathbf{RS}_{asym},\mathbf{RS}_{acyc},\mathbf{RS}_{order},\mathbf{CLA},\mathbf{OC}\}$ 
we denote by $r_M$ (``hit rate'') and $a_M$ (``area''), 
respectively, the proportions of human and synthetic subjects that are 
perfectly explained by $M$, and let 
\begin{eqnarray}
\label{selten1}  m^1_{_M} & := & r_{_M}-a_{_M}\\
\label{selten2} m^2_{_M} & := & \dfrac{r_{_M}}{a_{_M}}
\end{eqnarray}
be, respectively, that model's \textit{linear difference} and \textit{ratio} 
measures of predictive success at the given collection of choice problems 
\citep{selten91}. Both measures are increasing in $r$ and decreasing in $a$.
The linear measure---which is more frequently applied, and was axiomatized 
in \cite{selten91}---is bounded above by 1 and below by -1. 
Compared to the ratio measure,
it tends to favor models with better explanatory power (i.e. a higher $r_M$).
The ratio measure on the other hand is bounded below by 0 and unbounded above.
Because, by definition, it favors models with higher explanation/permissiveness
ratios, models which explain few human subjects' behavior
but are also unlikely to explain random choices could be ranked above 
models with the opposite empirical characteristics. 
The two measures are therefore complementary.

\begin{table}[!htbp]
	\centering
	\small
	\caption{\centering 
		Evaluating models according to Selten's (\citeyear{selten91})
		difference and ratio measures \linebreak of predictive success 
		(in parentheses, 95\% confidence intervals as in 
		\citeauthor{demuynck15}, \citeyear{demuynck15}).}
	\caption*{(a) Four Dresses.}
		\setlength{\tabcolsep}{5pt} 
		\renewcommand{\arraystretch}{1.1} 
		\begin{tabular}{|l|ccccc|}
			\hline
			& \multirow{2}{*}{\textbf{RS}$_{asym}$}
			& \multirow{2}{*}{\textbf{RS}$_{acyc}$}
			& \multirow{2}{*}{\textbf{RS}$_{order}$}
			& \multirow{2}{*}{\textbf{CLA}}
			& \multirow{2}{*}{\textbf{OC}} 
			\\
			&
			&
			&
			&
			&
			\\
			\hline
			$r$ 	& 0.0488 
			& 0.0488 
			& 0.0488 
			& 0.9146 
			& 0.6219 
			\\
			$a$ 	
			& 0.0119 
			& 0.0119
			& 0.0077
			& 0.3689
			& 0.3471
			\\
			$m^1$ 
			& 0.0369 
			& 0.0369 
			& 0.0411 
			& 0.5457 
			& 0.2748 
			\\
			& \footnotesize $(-0.002,0.076)$
			& \footnotesize $(-0.002,0.076)$
			& \footnotesize $(0.002,0.080)$
			& \footnotesize $(0.495,0.596)$
			& \footnotesize $(0.187,0.363)$
			\\
			$m^2$ 
			& 4.1008 
			& 4.1008 
			& 6.3376 
			& 2.4792 
			& 1.7917 
			\\
			& \footnotesize $(3.742,4.460)$
			& \footnotesize $(3.742,4.460)$
			& \footnotesize $(5.892,6.784)$
			& \footnotesize $(2.396,2.563)$
			& \footnotesize $(1.642,1.941)$
			\\
			\hline
		\end{tabular}
	\caption*{\ \\ (b) Five Ponchos.}
		\setlength{\tabcolsep}{5pt} 
		\renewcommand{\arraystretch}{1.1} 
		\begin{tabular}{|l|ccccc|}
			\hline
			& \multirow{2}{*}{\textbf{RS}$_{asym}$}
			& \multirow{2}{*}{\textbf{RS}$_{acyc}$}
			& \multirow{2}{*}{\textbf{RS}$_{order}$}
			& \multirow{2}{*}{\textbf{CLA}}
			& \multirow{2}{*}{\textbf{OC}}
			\\
			&
			&
			&
			&
			&
			\\
			\hline
			$r$ 	
			& 0.0958
			& 0.0958
			& 0.0898
			& 1.0000
			& 0.6227
			\\
			$a$ 	
			& 0.0030 
			& 0.0027 
			& 0.0016 
			& 1.0000 
			& 0.2094 
			\\
			$m^1$ 
			& 0.0928
			& 0.0931
			& 0.0882
			& 0.0000
			& 0.4133
			\\
			& \footnotesize $(0.055,0.130)$
			& \footnotesize $(0.056,0.131)$
			& \footnotesize $(0.052,0.125)$
			& \footnotesize $(0.000,0.000)$
			& \footnotesize $(0.352,0.475)$
			\\
			$m^2$ 
			& 31.933
			& 35.481
			& 56.125
			& 1.0000
			& 2.9737
			\\
			& \footnotesize $(31.249,32.617)$
			& \footnotesize $(34.760,36.202)$
			& \footnotesize $(55.215,57.035)$
			& \footnotesize $(1.000,1.000)$
			& \footnotesize $(2.839,3.109)$
			\\
			\hline
		\end{tabular}
	\caption*{\ \\ (c) Model rankings across the two datasets and measures.}
\begin{tabular}{|lr|lr||lr|lr|}
	\multicolumn{4}{c}{Four Dresses} & \multicolumn{4}{c}{Five Ponchos} \\
	\hline
	\multicolumn{2}{|c|}{$m^1$} 
	& \multicolumn{2}{c||}{$m^2$} 
	& \multicolumn{2}{c|}{$m^1$} 
	& \multicolumn{2}{c|}{$m^2$} \\
	\hline
	\textbf{CLA} & 0.545
	& \textbf{RS}$_{order}$ & 6.34
	& \textbf{OC} & 0.427
	& \textbf{RS}$_{order}$ & 56.12
	\\
	\hline
	\textbf{OC}	& 0.275
	& \textbf{RS}$_{acyc}$ & 4.10
	& \textbf{RS}$_{acyc}$ & 0.093
	& \textbf{RS}$_{acyc}$ & 35.48\\
	\hline
	\textbf{RS}$_{order}$ & 0.041
	& \textbf{RS}$_{asym}$ & 4.10
	& \textbf{RS}$_{asym}$ & 0.092
	& \textbf{RS}$_{asym}$ & 31.93\\
	\hline
	\textbf{RS}$_{acyc}$ & 0.037
	& \textbf{CLA} & 2.48
	& \textbf{RS}$_{order}$ & 0.088
	& \textbf{OC} & 2.973\\
	\hline
	\textbf{RS}$_{asym}$ & 0.037
	& \textbf{OC} & 1.79
	& \textbf{CLA} & 0.000
	& \textbf{CLA} & 1.000
	\\
	\hline
\end{tabular}
\label{tab:selten}
\end{table}

Indeed, this contrast between $m^1$ and $m^2$ is manifested in the results 
shown in Table \ref{tab:selten}, which were computed by applying our CDP 
models for each of the five choice models 
on 10,000 simulated datasets in each of the 4D and 5P data 
(the latter computations were completed in less than 24 hours in 
a regular laptop computer). 
In both experimental samples, $m^1$ puts one of the two limited-attention
models at the top whereas $m^2$ ranks \textbf{RS} first, with this model's
three constituent classes in reverse order of permissiveness
(namely, \textbf{RS}$_{order}$ first, followed by \textbf{RS}$_{acyc}$ 
and then \textbf{RS}$_{asym}$). 
Importantly, the top-ranked model under either measure 
and dataset is robustly so at the 5\% level of significance,
with confidence intervals computed as suggested in \cite{demuynck15}.\footnote{
Because all subjects in each dataset faced the same problems, the consistent 
estimator of the relevant variance-covariance matrix of $r$ and $a$ 
in \citet[p. 693]{demuynck15}, 
with respect to some model $M$ and given a sample of size $n$, 
has zero entries everywhere except for the variance of $r_{M,n}$. 
The latter is estimated by $v_{M,n}=r_{M,n}(1-r_{M,n})$
(note: we write $v_{M,n}$ instead of $v_{M,n,m}$, where 
$m$ here denotes the number of random datasets, due to the 
above-mentioned redundancies in the variance-covariance estimator). 
The 5\% confidence interval of measure $m^i$ for model $M$ 
given sample size $n$, denoted $m^i_{M,n}$ for $i=1,2$, 
is then determined by 
$\left[m^i_{M,n}-c_{5\%}\sqrt{\frac{v_{M,n}}{n}},
m^i_{M,n}+c_{5\%}\sqrt{\frac{v_{M,n}}{n}}\right]$, where $c_{5\%}\approx 1.645$
is the 5\% cutoff value of the standard normal cumulative 
density function.}

A result that is revealed by this analysis and may be worth highlighting 
is the fact that, despite the larger domain spanned by the 5P data
(5 items \& 16 menus, against 4 \& 11 in 4D), \textbf{CLA} explains perfectly
not only every human subject, but every synthetic random-behaving subject too.
This is despite that, for both \textbf{OC} and \textbf{CLA}, 
the consideration mapping of optimal solutions 
properly includes the choice made at some menu,
i.e. $\Gamma(A)\supsetneq C(A)$.\footnote{The \textbf{CLA} model's 
explanatory gains on both human and synthetic data gradually decrease 
as we expand this constraint by requiring to cover more menus. 
Introducing this extra analytical flexibility amounts to defining 
a distinct class of models which encompass the one introduced and 
axiomatically characterized in \cite{masatlioglu-nakajima-ozbay12}.
\cite{freer-nosratabadi-a} study a variation of \textbf{CLA} and \textbf{OC}
along those lines. There, however, ``attention floors'' (lower bounds on the 
number of alternatives that the decision maker pays attention to) apply across
all menus.}
An intuitive explanation for this finding
is that the 10 three-element menus from which choices were not observed 
in the 5P data give disproportionately more degrees of freedom in \textbf{CLA} 
compared to \textbf{OC} to define a rationalizing consideration mapping $\Gamma$
for any candidate matching preference relation.\footnote{Indeed, 
\eqref{AF-2} and \eqref{OC-2} differ meaningfully in this case: 
if $x$ is chosen at $\{v,w,x,y,z\}$ or in any of this menu's 4-element subsets,
\eqref{OC-2} implies that $x$ must necessarily belong to the consideration 
set $\Gamma(A)$ of every latent 3-element menu $A$ that contains $x$.
By not imposing this constraint, \eqref{AF-2} allows for more rationalization
possibilities under some $\Gamma$ for any candidate linear order $\succ$. 
For example, suppose $C(\{v,w,x,y\})=x$, $C(\{v,x\})=v$ and $C(\{x,y\})=y$
are part of a 5P choice dataset. 
To make the point clear while keeping other things equal, suppose 
$\Gamma^{OC}(\{v,w,x,y\})=\Gamma^{CLA}(\{v,w,x,y\})=\{w,x,y\}$.
Although $C(\{v,x,y\})$ is unobserved, we know from \eqref{OC-2} and 
the above data that $\Gamma^{OC}(\{v,x,y\})\supseteq\{x,y\}$ 
and, therefore, $\Gamma^{OC}(\{x,y\})=\{x,y\}$.
This implies, for example, that the candidate preference order 
$v\succ w \succ x \succ y$ could not explain these choices under \textbf{OC}. 
By contrast, \eqref{AF-2} allows $\Gamma^{CLA}$ to be agnostic on the latent 
menu $\{v,x,y\}$, and then to freely assign $\Gamma^{CLA}(\{v,x\})=\{v\}$,   
$\Gamma^{CLA}(\{x,y\})=\{y\}$. Such a $\Gamma^{CLA}$ mapping enables 
the observed choices to be in principle rationalizable 
by \textbf{CLA} under the above preference relation.}

\begin{figure}[!htbp]
	\centering
		\caption{\centering Undominated behavioral primitives that 
		explain subjects perfectly under one or more models.}
	\caption*{(a) Four Dresses \vspace{-15pt}}
	\includegraphics[width=0.49\linewidth]{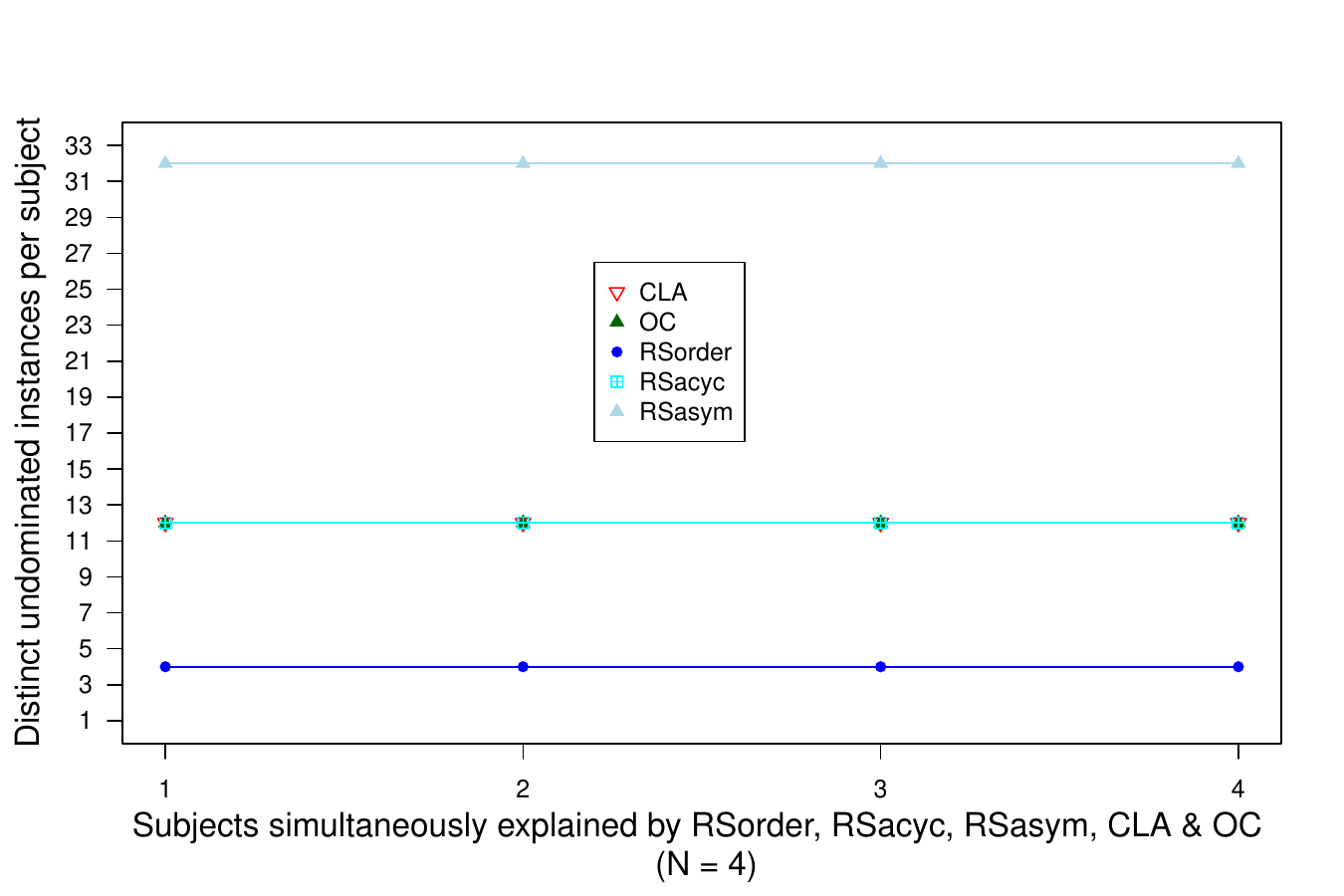}
	\includegraphics[width=0.49\linewidth]{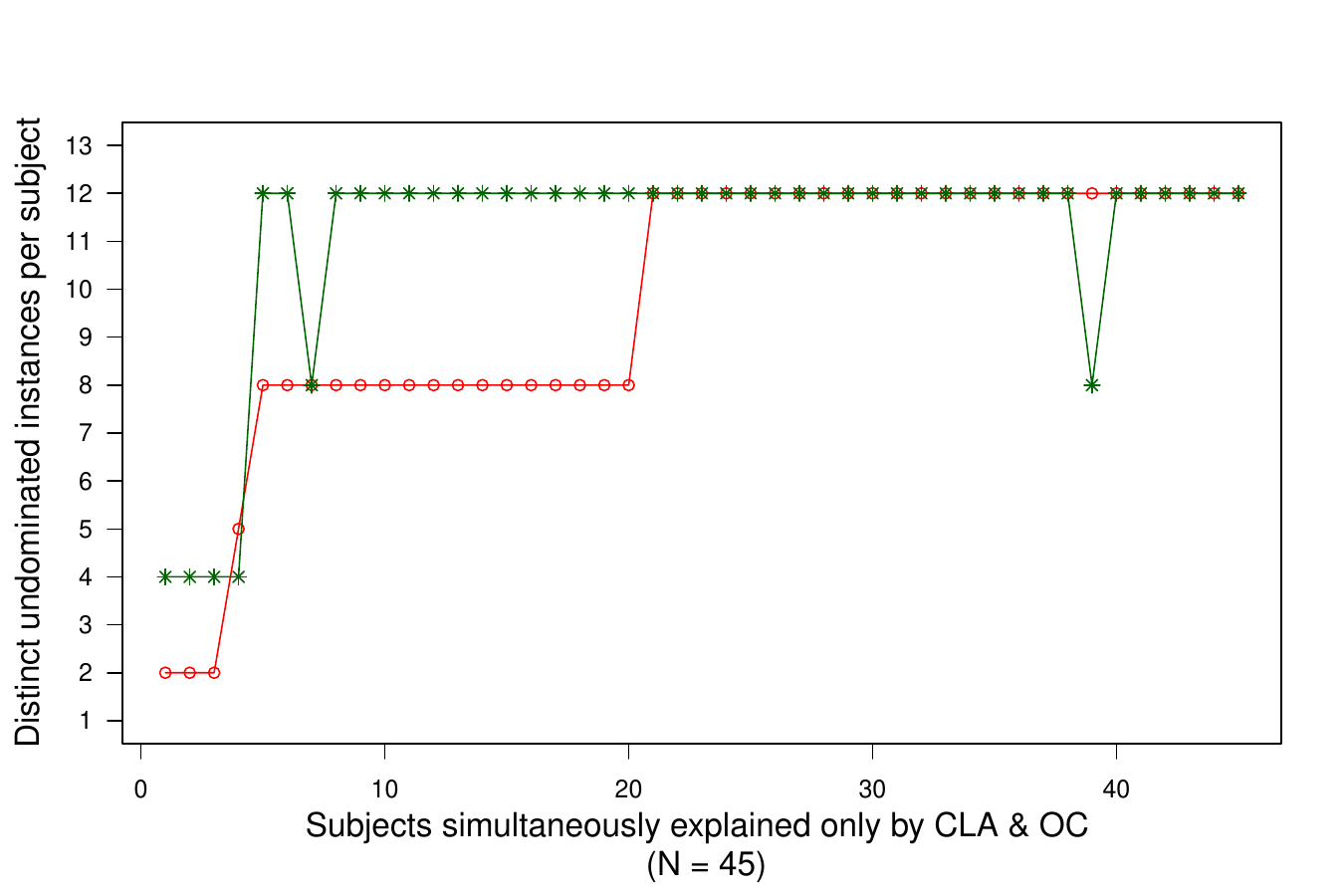}
	\includegraphics[width=0.49\linewidth]{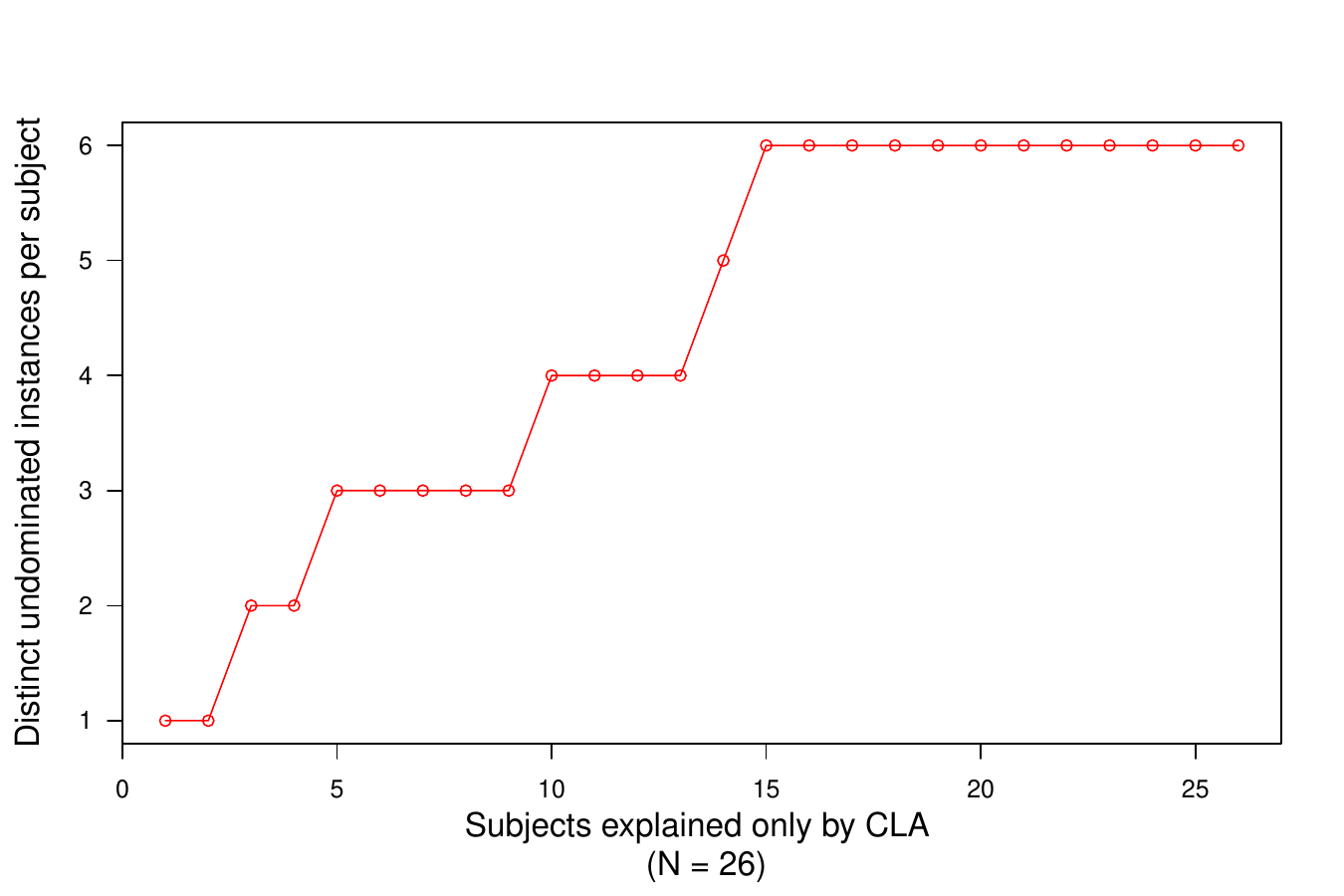}
	\includegraphics[width=0.49\linewidth]{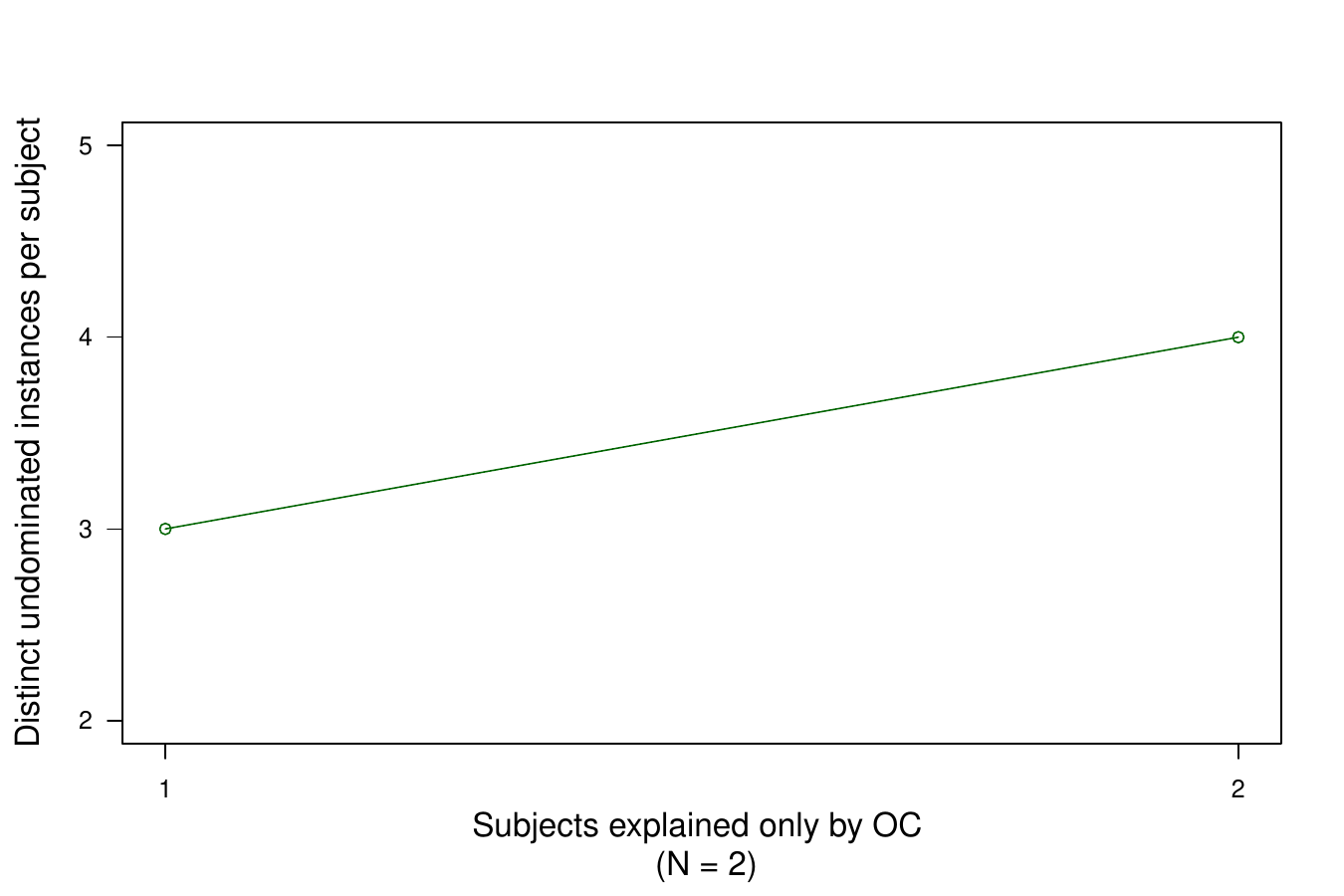}
	\caption*{(b) Five Ponchos \vspace{-15pt}}
	\includegraphics[width=0.49\linewidth]{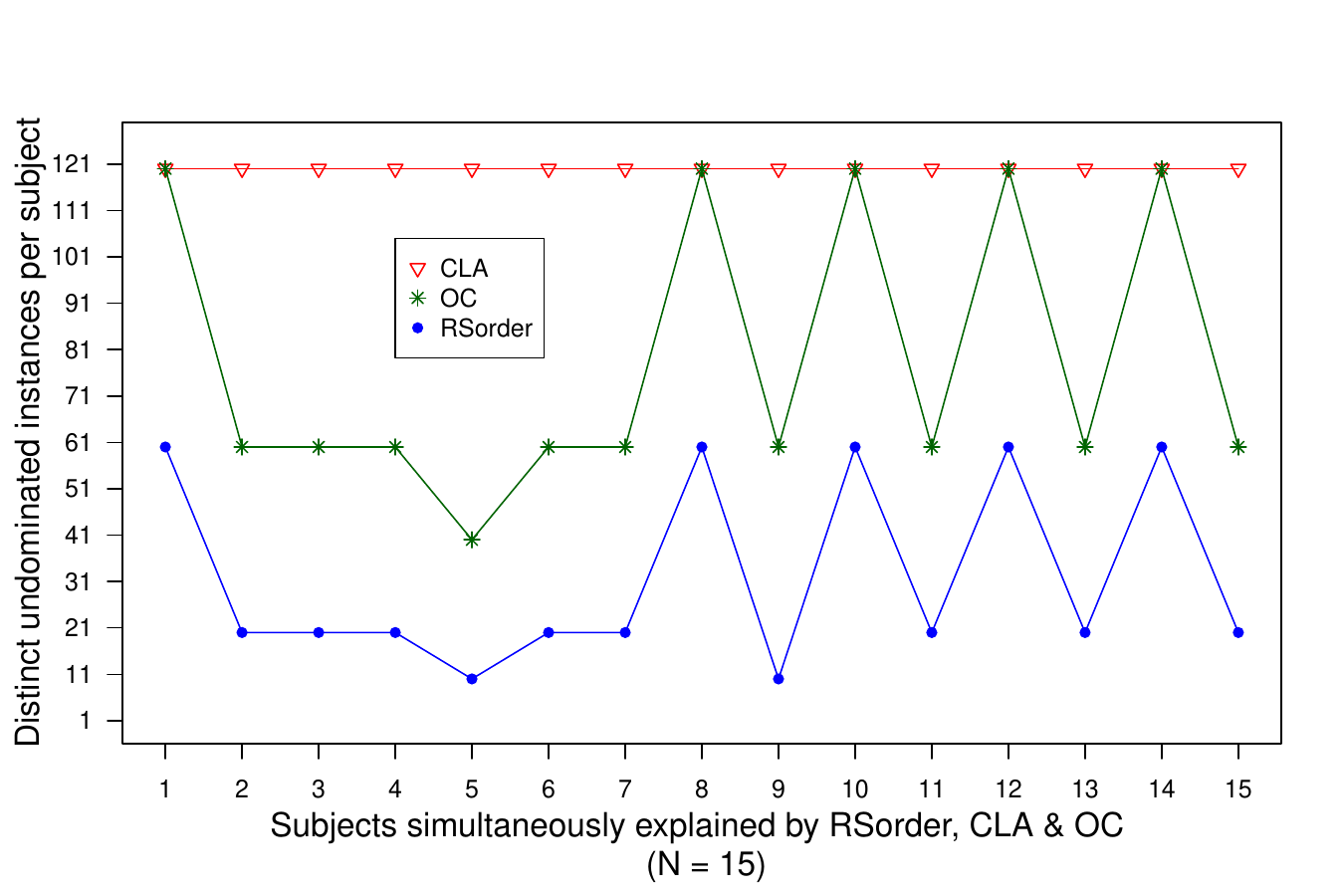}
	\includegraphics[width=0.49\linewidth]{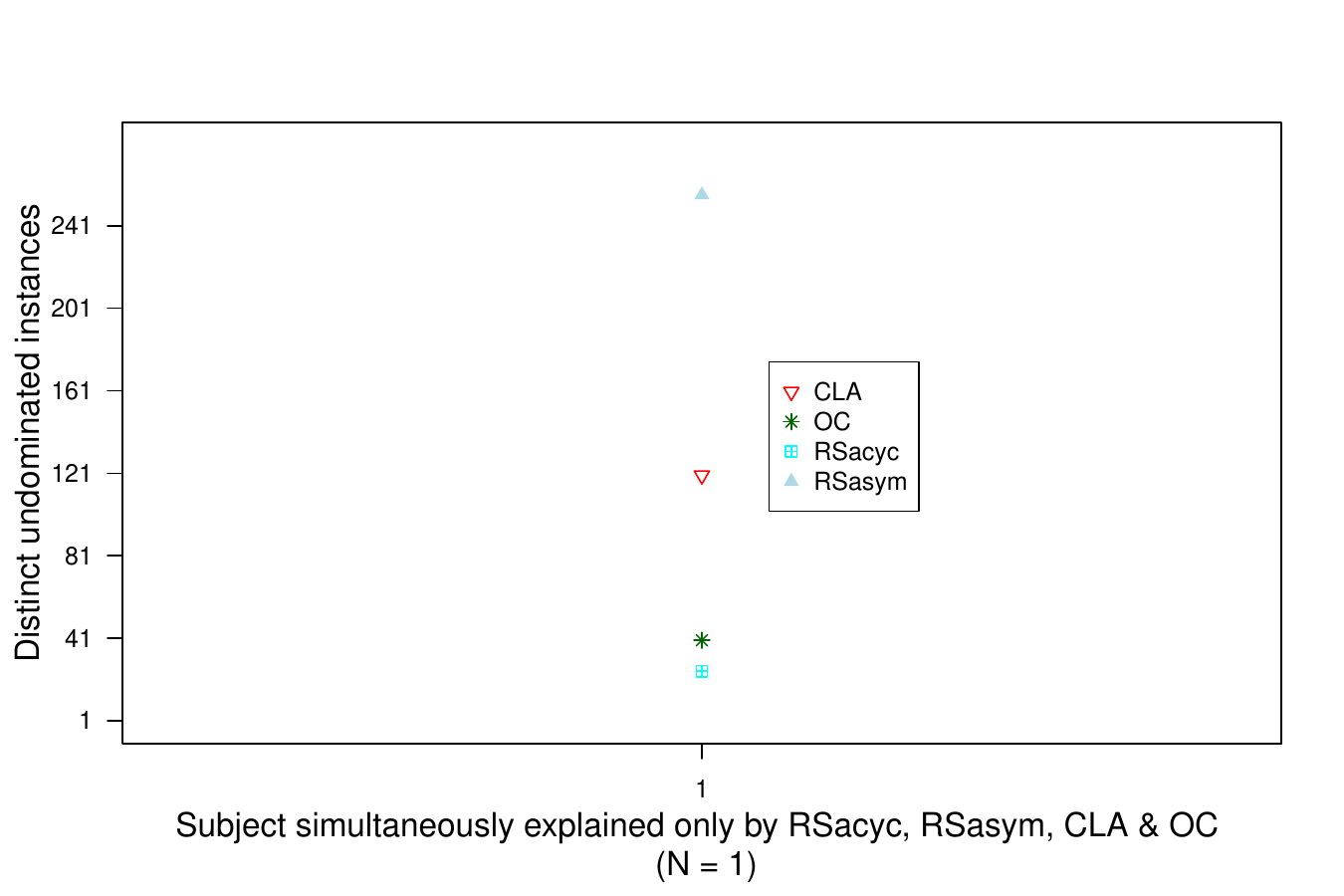}
	\includegraphics[width=0.49\linewidth]{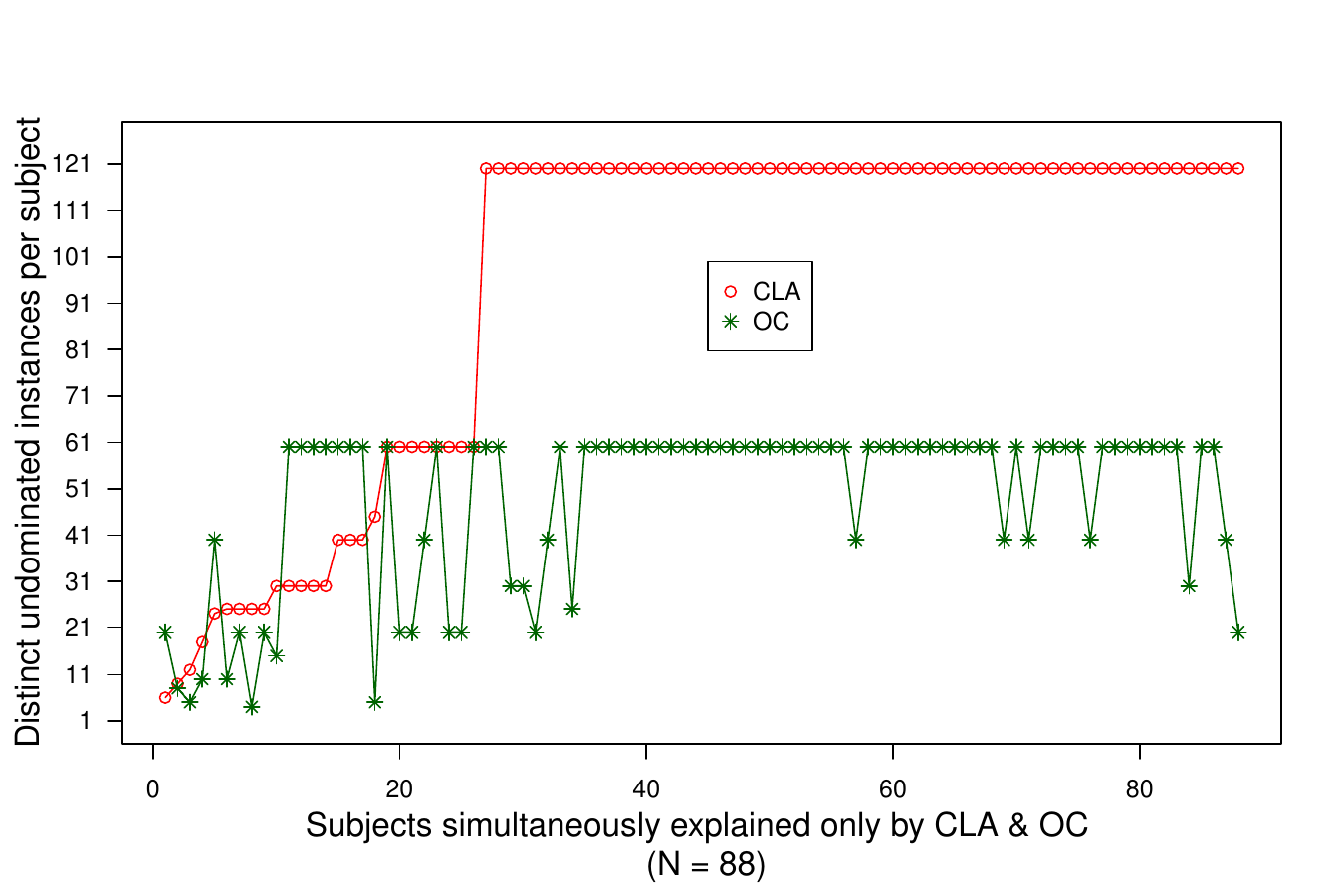}
	\includegraphics[width=0.49\linewidth]{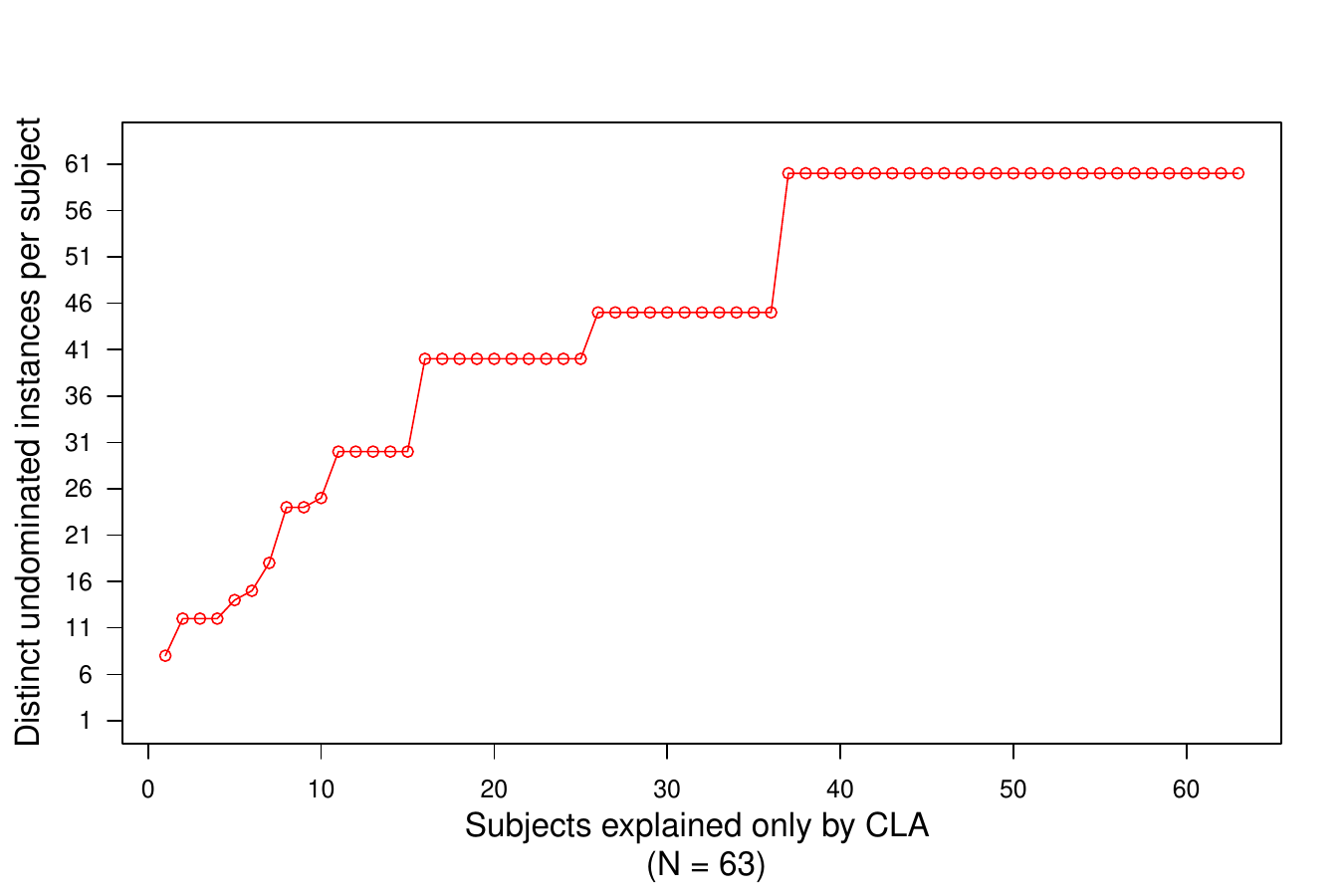}
	\label{fig:linecharts-LA}
\end{figure}

Next, taking the perfect fits of all three models and classes thereof
at face value, in Table \ref{tab:model-scores-filtered} 
and Figure \ref{fig:linecharts-LA} we summarize the results from 
enumerating compatible model primitives per subject 
after application of our CDP method, which eliminates
instances that are dominated according to the refinements of 
Definitions \ref{dfn:dominance-rs}
and \ref{dfn:dominance-la}.
In the 4D data we find that, for all 4 subjects 
who are perfectly explained by \textbf{RS}, 
there are 4, 12 and 32 undominated instances that are compatible 
with the \textbf{RS}$_{order}$, \textbf{RS}$_{acyc}$ and \textbf{RS}$_{asym}$ 
classes of the model, respectively. The average number of 
undominated primitives for subjects who are perfectly explained 
by \textbf{CLA} and \textbf{OC}---both of which also 
explain the above 4 subjects---is 
8 and 11 for \textbf{CLA} and \textbf{OC}, respectively. 
Importantly, however, despite these generally high numbers, for the 2 and 26 
subjects who are perfectly matched \textit{only} by \textbf{OC} and \textbf{CLA}, 
respectively, the compatible instances are markedly lower, down to 
3.5 and 4.4 on average (Figure \ref{fig:linecharts-LA}(a)). 
In fact, for two of the \textbf{CLA}-only subjects 
our analysis elicits the most welfare-informative 
possible fits by identifying a \textit{unique} compatible 
undominated primitive $(\Gamma,\succ)$ per subject. 
This demonstrates that the CDP approach can be extremely illuminating
in some cases.

For the 4D data we also enumerated 
the full sets of compatible primitives (i.e. those that also include
dominated ones) in order to quantify each model's
degree of non-identification prior to the application of 
any refinements but also---and more importantly---the degree
to which a model's compatible primitives are dominated. 
As Table \ref{tab:model-scores-filtered}(a) shows, 
such explanatory indeterminacy is highest in \textbf{CLA} 
(typically amounting to thousands of compatible primitives) 
and \textbf{RS}$_{asym}$, followed by \textbf{OC}, \textbf{RS}$_{acyc}$ and 
\textbf{RS}$_{order}$. 
What is arguably most striking here is the 
fact that between 83.3\% (\textbf{RS}$_{order}$) 
and 99.6\% (\textbf{CLA}) of all instances across all models 
are discarded as dominated.

While we did not enumerate the full list of compatible 
model-specific behavioral primitives before applying 
our dominance refinements in the 5P data,
we note that in the case of one subject 
this amounted to nearly 351,000 \textbf{CLA} primitives.
In these data we applied our CDP models and tool chain directly, 
as explained in Section \ref{subsec:cdp-primer}; in particular, we 
did not have to compute the full list of compatible primitives 
via our CP models first. 
We find that the average and median numbers 
of undominated primitives are higher for all models relative to their 
4D-data counterparts, with the most indeterminate again being 
\textbf{RS}$_{asym}$ (average: 512), which is 
now followed by \textbf{CLA} 
rather than \textbf{OC} (79 vs 53), and then by 
\textbf{RS}$_{acyc}$ (95) and \textbf{RS}$_{order}$ (32).
The generally higher indeterminacy here is caused by the larger number of 
alternatives and menus in the data, as well as by the absence of 
choices from menus with three elements. 
The model-specific average computation times in 5P also increased 
by a varying number of factors compared to 4D 
($\approx$782 for \textbf{RS}$_{asym}$ vs. $\approx$4 for \textbf{CLA}).
However, all CDP computations---both for model scores and for the 
enumeration of undominated instances---were completed in a few hours
on a regular laptop.

\begin{table}[!htbp]
	\caption{\centering 
		Average and median undominated model primitives and  
		computation times, and the corresponding statistics 
		for approximately compatible 
		fits under Rational Choice.}
	\caption*{(a) Four Dresses}
	\small
	\makebox[\linewidth][c]{
		\setlength{\tabcolsep}{5pt} 
		\renewcommand{\arraystretch}{1.1} 
		\begin{tabular}{|l|c|c|c|c|c|c|}
			\hline
			&\textbf{RS$_{order}$}
			&\textbf{RS$_{acyc}$}
			&\textbf{RS$_{asym}$}
			&\textbf{CLA}
			&\textbf{OC}
			&\textbf{RC}\\
			\hline
			Compatible subjects
			& 5\% 
			& 5\% 
			& 5\% 
			& 91.5\% 
			& 62\% 
			& (100\%) 
			\\
			\hline
			Mean (median) primitives 
			& \multirow{2}{*}{24}
			& \multirow{2}{*}{480}
			& \multirow{2}{*}{768}
			& \multirow{2}{*}{1932 (1643)}
			& \multirow{2}{*}{664 (638)}
			& \multirow{4}{*}{1.51 (1)} \\
			per matching model
			&
			&
			&
			&
			& 
			&\\
			\cline{1-6}
			Mean (median) undominated 
			& \multirow{2}{*}{4 ($\cdot$)}
			& \multirow{2}{*}{12 ($\cdot$)}
			& \multirow{2}{*}{32 ($\cdot$)}
			& \multirow{2}{*}{8.01 (8)}
			& \multirow{2}{*}{10.9 (12)}
			&  \\
			primitives per matching model
			&
			&
			&
			&
			&
			& \\
			\hline
			Mean (median) total 
			& \multirow{2}{*}{0.0001 ($\cdot$)}
			& \multirow{2}{*}{0.003 ($\cdot$)}
			& \multirow{2}{*}{0.007 ($\cdot$)}
			& \multirow{2}{*}{0.078 (0.075)}
			& \multirow{2}{*}{0.081 (0.079)}
			& \multirow{2}{*}{0.0007 (0.0002)} \\
			computation time, in seconds
			& 
			& 
			& 
			& 
			& 
			& \\
			\hline
		\end{tabular}
	}
	\caption*{\ \\ (b) Five Ponchos}
\small
\makebox[\linewidth][c]{
	\setlength{\tabcolsep}{5pt} 
	\renewcommand{\arraystretch}{1.1} 
	\begin{tabular}{|l|c|c|c|c|c|c|}
		\hline
		&\textbf{RS$_{order}$}
		&\textbf{RS$_{acyc}$}
		&\textbf{RS$_{asym}$}
		&\textbf{CLA}
		&\textbf{OC}
		&\textbf{RC}\\
		\hline
		Compatible subjects
		& 9\% 
		& 9.5\% 
		& 9.5\% 
		& 100\% 
		& 62\% 
		& (100\%) 
		\\
		\hline
		Mean (median) undominated 
		& \multirow{2}{*}{32 (20)}
		& \multirow{2}{*}{94.7 (60)}
		& \multirow{2}{*}{640 (512)}
		& \multirow{2}{*}{78.8 (60)}
		& \multirow{2}{*}{52.8 (60)}
		& \multirow{2}{*}{1.57 (1)} \\
		primitives per matching model
		&
		&
		&
		&
		&
		& \\
		\hline
		Mean (median) total 
		& \multirow{2}{*}{0.053 (0.049)}
		& \multirow{2}{*}{0.142 (0.097)}
		& \multirow{2}{*}{5.48 (3.05)}
		& \multirow{2}{*}{0.295 (0.23)}
		& \multirow{2}{*}{0.747 (0.794)}
		& \multirow{2}{*}{0.038 ($\cdot$)} \\
		computation time, in seconds
		& 
		& 
		& 
		& 
		& 
		& \\
		\hline
		
	\end{tabular}
}
	\label{tab:model-scores-filtered}
	\footnotesize \ \\
	\centerline{Note: 
	$(\cdot)$ means that the median coincides with the mean.}
\end{table}

\begin{table}[!htbp]
	\centering
	\caption{\centering
		Percentage of subjects whose perfect or approximate model fits \linebreak
		were unlikely under random behavior (Bronars, \citeyear{bronars87})*.}
	\subtable[Four Dresses]{	
		\setlength{\tabcolsep}{5pt} 
		\renewcommand{\arraystretch}{1.1} 
		\begin{tabular}{|l|cc|}
		\hline
		\textbf{RC}			& 68.3\%& (2) \\	
		\textbf{RS}$_{order}$& 5\% 	& (1) \\	
		\textbf{RS}$_{acyc}$	& 5\% 	& (1)\\	
		\textbf{RS}$_{asym}$	& 5\% 	& (1)\\	
		\textbf{CLA} 		& 0\% 	& (0)\\	
		\textbf{OC}			& 0\% 	& (0)\\	
		\hline
	\end{tabular}
}
\subtable[Five Ponchos]{	
	\setlength{\tabcolsep}{5pt} 
	\renewcommand{\arraystretch}{1.1} 
	\begin{tabular}{|l|cc|}
	\hline
	\textbf{RC}			& 79.6\%& (3) \\ 
	\textbf{RS}$_{order}$& 64.6\%& (2) \\ 
	\textbf{RS}$_{acyc}$	& 9.6\% & (1) \\
	\textbf{RS}$_{asym}$	& 9.6\% & (1) \\
	\textbf{CLA} 		& 0\% 	& (0) \\
	\textbf{OC}			& 0\% 	& (0) \\
	\hline
\end{tabular}
}	\label{tab:bronars}

{\footnotesize*In parenthesis are the 2.5th percentile 
model scores derived from simulations.}
\end{table}

These results highlight the potential of the prominent 
shorlisting and limited-attention 
theories of bounded-rational choice, in conjunction with sophisticated 
theory-based computational methods such as the C(D)P ones 
that we propose, to extract non-trivial behavioral insights from
human choices that deviate from the rational model. 
Yet the general picture that emerges from the preceding analysis 
generally points toward a considerable amount of 
indeterminacy still remaining for each of these models after application
of the relevant dominance-based criteria 
for selecting welfare-relevant model primitives . 
Perhaps a natural benchmark relative to which the degree of this indeterminacy
might be assessed is how \textbf{RC} performs in this respect on 
the same data. Indeed, although the rational-choice model is fully identified
when choices from all possible menus are available (as is the case here), 
since our empirical analysis is limited to subjects who can only be 
\textit{approximately} explained by \textbf{RC}, there is no \textit{a priori} 
guarantee that any such approximate fits will also be generated by a single 
preference ordering per subject. Yet this is indeed what we find: 
an average (median) number of 1.51 (1) and 1.57 (1) strict preference relations
per subject in the 4D and 5P data, respectively, 
almost always corresponding to approximate fits with a distance score of 1 or 2.
In other words, with the exception of few subjects, 
the \textit{approximate} \textbf{RC} fit generally gives more
focused welfare-relevant information than the \textit{perfect} fit of 
every bounded-rational model under study. 

This finding, in turn, raises the question: 
Is it more appropriate for such human decision makers to be viewed
as if they behaved perfectly in the ways described by \textbf{CLA}, 
\textbf{OC} and \textbf{RS}, or, instead, as ``noisy'' 
utility maximizers?\footnote{A distinct related question is: 
How should an analyst decide which one of several bounded-rational 
models provides the best way to think about a choice dataset 
that is explained by all of them and may also be (in)compatible 
with \textbf{RC}? In particular, what is the role of non-choice data 
in such discerning attempts? We refer the reader to discussions in 
\cite{caplin-shotter08,manzini-mariotti14,declippel-rozen24} for more on this.}
While there may be little hope to arrive at a definitive and generally 
applicable answer, using synthetic data featuring random behavior in the way
originally suggested in \cite{bronars87} could be helpful in that direction.
More specifically, we can use the model-specific distributions 
of distance scores on the \textit{synthetic} datasets (which are available
from the predictive-success analysis that was presented earlier) and, by analogy 
to the two-tailed statistical test at the 5\% significance level, determine the 
cut-off scores that correspond to the 2.5th percentile values. 
Finally, once these are pinned down 
we can discard those \textit{human} datasets whose model-specific scores 
weakly exceed the respective cut-off values, on the grounds that we cannot 
plausibly rule out that the (perfect or approximate) fits they give rise to
could not have been achieved by a person making choices at random.

As is detailed in Table \ref{tab:bronars}, the majority of approximate fits 
that are achieved by \textbf{RC} survive this robustness check in both the 
4D and 5P experimental data. Furthermore, all perfect fits of all three 
\textbf{RS} model classes also survive, while the opposite is true for 
\textbf{CLA} and \textbf{OC} (see also the high $a_M$ values 
in Table \ref{tab:selten}).
Finally, this analysis suggests that the almost perfect fit achieved by 
the most structured version of the shortlisting model, \textbf{RS}$_{order}$, 
for 93 of 167 subjects in the 5P sample should also be considered
acceptable according to that criterion. 
Although this method is useful in narrowing down 
the imperfect rational and perfect bounded-rational explanations that 
are simultaneously possible for a given subject, we note that it does not 
eliminate the need for additional scrutiny and discretion to be 
exercised by the analyst interested in arriving at a single best explanation.

\section{Concluding Remarks}

Many deterministic models of bounded-rational general choice 
have been developed in the past two decades 
with the overarching goal of explaining in introspectively 
intuitive ways behavioral phenomena that the textbook model of
rational choice is unable to. 
The extensive theoretical interest in this direction, however, 
has not been met by an analogous interest in, and advances on, 
the rigorous empirical testing of these models in the kinds of general, 
riskless choice environments that motivated their development. 
The high computational complexity associated with in-depth analyses 
of these models is arguably one explanation for the relative 
absence of such work.\footnote{It is also one of the reasons 
for the increased theoretical interest in the formulation of 
bounded-rational \textit{random} choice models more recently. 
Under distinct assumptions on the nature of 
the available data, and under additional identifying assumptions, 
these models have proved more amenable to empirical testing with 
econometric methods 
\citep{cattaneo-ma-masatlioglu-suleymanov20,
dardanoni-manzini-mariotti-petri-tyson23,filiz-ozbay-masatlioglu23}.} 
In the present study we tackle this problem directly by introducing into the 
bounded-rationality and revealed-preference literatures 
constraint and constraint-dominance 
programming methods, and a comprehensive, 
free and open-source such tool chain from the frontier 
of research at the intersection of discrete optimization, computer science,
artificial intelligence and operations research. 
This facilitates detailed---and in full generality---non-parametric 
goodness-of-fit tests of what are arguably two of the most prominent 
classes of such models, and assessing how informative such 
fits are to any analyst interested to use them toward inferring 
decision makers' welfare-relevant information---namely, 
their shortlisting criteria or attention constraints and preferences---in practical 
choice situations.

The illustrative empirical application of our methods and tools 
on the two experimental datasets with riskless choices over 
four and five alternatives that we considered shows that 
limited-attention models explain better than 
their shortlisting counterparts, but are also more permissive.
In addition, of all the models and special classes thereof that we 
analyzed, only shortlisting with transitive rationales
achieves (perfect or imperfect) empirical fits that pass the 
``non-randomness'' robustness test. This structured special
case of the shortlisting model is also the one that, in both 
datasets, generates the most parsimonious fits ---namely those 
with the smallest number of compatible behavioral primitives.
Despite their higher explanatory ability, however, all 
bounded-rational choice models that we studied fared worse 
in terms of parsimony than the textbook model of rational choice, whose 
approximate empirical fits were typically associated with 
a unique compatible primitive/preference orderings in these data.

Concluding, we hope that this paper's methodological contribution 
and empirical findings will aid the development of new 
bounded-rational models of general choice 
in which the---now measurable---degrees of permissiveness and 
explanatory indeterminacy will be evaluated alongside the 
models' computational complexity and 
ability to explain observed behavioral phenomena.

\bibliographystyle{ecta}
\bibliography{BRCP}

\begin{thebibliography}{66}
\newcommand{\enquote}[1]{``#1''}
\expandafter\ifx\csname natexlab\endcsname\relax\def\natexlab#1{#1}\fi

\bibitem[\protect\citeauthoryear{Akg{\"u}n, Frisch, Gent, Jefferson, Miguel,
  and Nightingale}{Akg{\"u}n et~al.}{2022}]{akgun2022conjure}
\textsc{Akg{\"u}n, {\"O}., A.~M. Frisch, I.~P. Gent, C.~Jefferson, I.~Miguel,
  and P.~Nightingale} (2022): \enquote{Conjure: Automatic Generation of
  Constraint Models from Problem Specifications,} \emph{Artificial
  Intelligence}, 310, 103751.

\bibitem[\protect\citeauthoryear{Andrews, Fudenberg, Lei, Liang, and
  Wu}{Andrews et~al.}{2025}]{andrews-fudenberg-lei-liang-wu}
\textsc{Andrews, I., D.~Fudenberg, L.~Lei, A.~Liang, and C.~Wu} (2025):
  \enquote{The Transfer Performance of Economic Models,} \emph{Working Paper}.

\bibitem[\protect\citeauthoryear{Apesteguia and Ballester}{Apesteguia and
  Ballester}{2015}]{apesteguia-ballester15}
\textsc{Apesteguia, J. and M.~Ballester} (2015): \enquote{A Measure of
  Rationality and Welfare,} \emph{Journal of Political Economy}, 123,
  1278--1310.

\bibitem[\protect\citeauthoryear{Apt}{Apt}{2003}]{apt03}
\textsc{Apt, K.~R.} (2003): \emph{Principles of Constraint Programming},
  Cambridge University Press.

\bibitem[\protect\citeauthoryear{Au and Kawai}{Au and Kawai}{2011}]{au-kawai11}
\textsc{Au, P.~H. and K.~Kawai} (2011): \enquote{Sequentially Rationalizable
  Choice with Transitive Rationales,} \emph{Games and Economic Behavior}, 73,
  608--614.

\bibitem[\protect\citeauthoryear{Beatty and Crawford}{Beatty and
  Crawford}{2011}]{beatty-crawford11}
\textsc{Beatty, T. K.~M. and I.~A. Crawford} (2011): \enquote{How Demanding is
  the Revealed Preference Approach to Demand?} \emph{American Economic Review},
  101, 2782--2795.

\bibitem[\protect\citeauthoryear{Bernheim and Rangel}{Bernheim and
  Rangel}{2009}]{bernheim-rangel09}
\textsc{Bernheim, B.~D. and A.~Rangel} (2009): \enquote{Beyond Revealed
  Preference: Choice-Theoretic Foundations for Behavioral Welfare Economics,}
  \emph{Quarterly Journal of Economics}, 124, 51--104.

\bibitem[\protect\citeauthoryear{Bessiere}{Bessiere}{2006}]{bessiere06}
\textsc{Bessiere, C.} (2006): \enquote{Constraint Propagation,} in
  \emph{Handbook of Constraint Programming}, ed. by F.~Rossi, P.~van Beek, and
  T.~Walsh, Elsevier, chap.~3.

\bibitem[\protect\citeauthoryear{Bronars}{Bronars}{1987}]{bronars87}
\textsc{Bronars, S.~G.} (1987): \enquote{The Power of Non-parametric Tests of
  Preference Maximization,} \emph{Econometrica}, 55, 693--698.

\bibitem[\protect\citeauthoryear{Caplin, Dean, and Martin}{Caplin
  et~al.}{2011}]{caplin-dean-martin11}
\textsc{Caplin, A., M.~Dean, and D.~Martin} (2011): \enquote{Search and
  Satisficing,} \emph{American Economic Review}, 101, 2899--2922.

\bibitem[\protect\citeauthoryear{Caplin and Schotter}{Caplin and
  Schotter}{2008}]{caplin-shotter08}
\textsc{Caplin, A. and A.~Schotter}, eds. (2008): \emph{The Foundations of
  Positive and Normative Economics: A Handbook}, New York: Oxford University
  Press.

\bibitem[\protect\citeauthoryear{Cattaneo, Ma, Masatlioglu, and
  Suleymanov}{Cattaneo et~al.}{2020}]{cattaneo-ma-masatlioglu-suleymanov20}
\textsc{Cattaneo, M.~D., X.~Ma, Y.~Masatlioglu, and E.~Suleymanov} (2020):
  \enquote{A Random Attention Model,} \emph{Journal of Political Economy}, 128,
  2796--2836.

\bibitem[\protect\citeauthoryear{Cherchye, Demuynck, {d}e {R}ock, and
  Lanier}{Cherchye et~al.}{2025}]{cherchye-demuynck-derock-lanier25}
\textsc{Cherchye, L., T.~Demuynck, B.~{d}e {R}ock, and J.~Lanier} (2025):
  \enquote{Are Consumers (Approximately) Rational?: Shifting the Burden of
  Proof,} \emph{The Review of Economics and Statistics}, 107, 1652--1666.

\bibitem[\protect\citeauthoryear{Choi, Fisman, Gale, and Kariv}{Choi
  et~al.}{2007}]{choi-fisman-gale-kariv07}
\textsc{Choi, S., R.~Fisman, D.~Gale, and S.~Kariv} (2007):
  \enquote{Consistency and Heterogeneity of Individual Behavior under
  Uncertainty,} \emph{American Economic Review}, 97, 1921--1938.

\bibitem[\protect\citeauthoryear{Chu and Stuckey}{Chu and
  Stuckey}{2015}]{chu-stuckey15}
\textsc{Chu, G. and P.~J. Stuckey} (2015): \enquote{Dominance Breaking
  Constraints,} \emph{Constraints}, 20, 155--182.

\bibitem[\protect\citeauthoryear{Costa-Gomes, Cueva, Gerasimou, and
  Tejiščák}{Costa-Gomes et~al.}{2022}]{CCGT22}
\textsc{Costa-Gomes, M., C.~Cueva, G.~Gerasimou, and M.~Tejiščák} (2022):
  \enquote{Choice, Deferral and Consistency,} \emph{Quantitative Economics},
  13, 1297--1318.

\bibitem[\protect\citeauthoryear{Dardanoni, Manzini, Mariotti, Petri, and
  Tyson}{Dardanoni et~al.}{2023}]{dardanoni-manzini-mariotti-petri-tyson23}
\textsc{Dardanoni, V., P.~Manzini, M.~Mariotti, H.~Petri, and C.~J. Tyson}
  (2023): \enquote{Mixture Choice Data: Revealing Preferences and Cognition,}
  \emph{Journal of Political Economy}, 131, 687--715.

\bibitem[\protect\citeauthoryear{de~Clippel and Rozen}{de~Clippel and
  Rozen}{2021}]{declippel-rozen21}
\textsc{de~Clippel, G. and K.~Rozen} (2021): \enquote{Bounded Rationality and
  Limited Datasets,} \emph{Theoretical Economics}, 16, 359--380.

\bibitem[\protect\citeauthoryear{de~Clippel and Rozen}{de~Clippel and
  Rozen}{2024}]{declippel-rozen24}
---\hspace{-.1pt}---\hspace{-.1pt}--- (2024): \enquote{Bounded Rationality in
  Choice Theory: A Survey,} \emph{Journal of Economic Literature}, 62,
  995--1039.

\bibitem[\protect\citeauthoryear{Dean and Martin}{Dean and
  Martin}{2016}]{dean-martin16}
\textsc{Dean, M. and D.~Martin} (2016): \enquote{Measuring Rationality with the
  Minimum Cost of Revealed Preference Violations,} \emph{The Review of
  Economics and Statistics}, 98, 524--534.

\bibitem[\protect\citeauthoryear{Dechter}{Dechter}{2003}]{dechter03}
\textsc{Dechter, R.} (2003): \emph{Constraint Processing}, Morgan Kaufmann.

\bibitem[\protect\citeauthoryear{Demetry and Hjertstrand}{Demetry and
  Hjertstrand}{2023}]{demetry-hjertstrand23}
\textsc{Demetry, M. and P.~Hjertstrand} (2023): \enquote{Consistent Subsets:
  Computing the Houtman–Maks Index in {S}tata,} \emph{The Stata Journal}, 23,
  578--588.

\bibitem[\protect\citeauthoryear{Demuynck}{Demuynck}{2011}]{demuynck11}
\textsc{Demuynck, T.} (2011): \enquote{The Computational Complexity of
  Rationalizing Boundedly Rational Choice Behavior,} \emph{Journal of
  Mathematical Economics}, 47, 425--433.

\bibitem[\protect\citeauthoryear{Demuynck}{Demuynck}{2015}]{demuynck15}
---\hspace{-.1pt}---\hspace{-.1pt}--- (2015): \enquote{Statistical Inference
  for Measures of Predictive Success,} \emph{Theory and Decision}, 79,
  689--699.

\bibitem[\protect\citeauthoryear{Demuynck and Rehbeck}{Demuynck and
  Rehbeck}{2023}]{demuynck-rehbeck23}
\textsc{Demuynck, T. and J.~Rehbeck} (2023): \enquote{Computing Revealed
  Preference Goodness-of-Fit Measures with Integer Programming,} \emph{Economic
  Theory}, 76, 1175--1195.

\bibitem[\protect\citeauthoryear{Dutta and Horan}{Dutta and
  Horan}{2015}]{dutta-horan15}
\textsc{Dutta, R. and S.~Horan} (2015): \enquote{Inferring Rationales from
  Choice,} \emph{American Economic Journal: Microeconomics}, 7, 179--201.

\bibitem[\protect\citeauthoryear{Ellis and Freeman}{Ellis and
  Freeman}{2024}]{ellis-freeman24}
\textsc{Ellis, A. and D.~J. Freeman} (2024): \enquote{Revealing Choice
  Bracketing,} \emph{American Economic Review}, 114, 2668--2700.

\bibitem[\protect\citeauthoryear{Filiz-Ozbay and Masatlioglu}{Filiz-Ozbay and
  Masatlioglu}{2023}]{filiz-ozbay-masatlioglu23}
\textsc{Filiz-Ozbay, E. and Y.~Masatlioglu} (2023): \enquote{Progressive Random
  Choice,} \emph{Journal of Political Economy}, 131, 716--750.

\bibitem[\protect\citeauthoryear{Fr\'{e}chette, Vespa, and
  Yuksel}{Fr\'{e}chette et~al.}{2026}]{frechette-vespa-yuksel26}
\textsc{Fr\'{e}chette, G., E.~Vespa, and S.~Yuksel} (2026): \enquote{People as
  Intuitive Modelers: How Model Complexity Adapts to Data,} \emph{Working
  Paper}.

\bibitem[\protect\citeauthoryear{Freer and Nosratabadi}{Freer and
  Nosratabadi}{2024}]{freer-nosratabadi-b}
\textsc{Freer, M. and H.~Nosratabadi} (2024): \enquote{On the Welfare
  (Ir)Relevance of Two-Stage Models,} \emph{arXiv:2411.08263}.

\bibitem[\protect\citeauthoryear{Freer and Nosratabadi}{Freer and
  Nosratabadi}{2026}]{freer-nosratabadi-a}
---\hspace{-.1pt}---\hspace{-.1pt}--- (2026): \enquote{Revealed Preference
  Analysis Under Limited Attention,} \emph{Journal of Economic Behavior \&
  Organization}, 246, 107568.

\bibitem[\protect\citeauthoryear{Frisch, Harvey, Jefferson,
  Mart{\'\i}nez-Hern{\'a}ndez, and Miguel}{Frisch
  et~al.}{2008}]{frisch2008essence}
\textsc{Frisch, A.~M., W.~Harvey, C.~Jefferson, B.~Mart{\'\i}nez-Hern{\'a}ndez,
  and I.~Miguel} (2008): \enquote{Essence: A Constraint Language for Specifying
  Combinatorial Problems,} \emph{Constraints}, 13, 268--306.

\bibitem[\protect\citeauthoryear{Fudenberg, Gao, and You}{Fudenberg
  et~al.}{2026{\natexlab{a}}}]{fudenberg-gao-you}
\textsc{Fudenberg, D., W.~Y. Gao, and Z.~You} (2026{\natexlab{a}}):
  \enquote{Model Restrictiveness in Functional and Structural Settings,}
  \emph{arXiv Working Paper 2602.07688}.

\bibitem[\protect\citeauthoryear{Fudenberg, Kleinberg, Liang, and
  Mullainathan}{Fudenberg
  et~al.}{2022}]{fudenberg-kleinberg-liang-muillainathan}
\textsc{Fudenberg, D., J.~Kleinberg, A.~Liang, and S.~Mullainathan} (2022):
  \enquote{Measuring the Completeness of Economic Models,} \emph{Journal of
  Political Economy}, 130, 956--990.

\bibitem[\protect\citeauthoryear{Fudenberg, Liang, and Gao}{Fudenberg
  et~al.}{2026{\natexlab{b}}}]{fudenberg-lian-gao21}
\textsc{Fudenberg, D., A.~Liang, and W.~Gao} (2026{\natexlab{b}}): \enquote{How
  Flexible is that Functional Form? Quantifying the Restrictiveness of
  Theories,} \emph{The Review of Economics and Statistics}, 108, 194--209.

\bibitem[\protect\citeauthoryear{Gent, Jefferson, and Miguel}{Gent
  et~al.}{2006}]{MINION}
\textsc{Gent, I.~P., C.~Jefferson, and I.~Miguel} (2006): \enquote{{Minion:} A
  Fast, Scalable, Constraint Solver,} in \emph{Proceedings of the 17th European
  Conference on Artificial Intelligence}, IOS Press.

\bibitem[\protect\citeauthoryear{Gerasimou}{Gerasimou}{2026}]{gerasimou26}
\textsc{Gerasimou, G.} (2026): \enquote{Eliciting and Distinguishing Between
  Weak and Incomplete Preferences: Theory, Experiment and Computation,}
  \emph{Games and Economic Behavior}, 158, 737--762.

\bibitem[\protect\citeauthoryear{Gerasimou and Tejiščák}{Gerasimou and
  Tejiščák}{2018}]{gerasimou-tejiscak-prest}
\textsc{Gerasimou, G. and M.~Tejiščák} (2018): \enquote{Prest: Open-Source
  Software for Computational Revealed Preference Analysis,} \emph{Journal of
  Open Source Software}, 3(30), 1015.

\bibitem[\protect\citeauthoryear{Guns, Stuckey, and Tack}{Guns
  et~al.}{2018}]{guns-stuckey-tack18}
\textsc{Guns, T., P.~J. Stuckey, and G.~Tack} (2018): \enquote{Solution
  Dominance over Constraint Satisfaction Problems,} \emph{CoRR},
  abs/1812.09207.

\bibitem[\protect\citeauthoryear{Heufer and Hjertstrand}{Heufer and
  Hjertstrand}{2015}]{heufer-hjertstrand15}
\textsc{Heufer, J. and P.~Hjertstrand} (2015): \enquote{Consistent Subsets:
  Computationally Feasible Methods to Compute the Houtman-Maks Index,}
  \emph{Economics Letters}, 128.

\bibitem[\protect\citeauthoryear{Houtman and Maks}{Houtman and
  Maks}{1985}]{houtman-maks85}
\textsc{Houtman, M. and J.~A. Maks} (1985): \enquote{Determining All Maximal
  Data Subsets Consistent with Revealed Preference,} \emph{Kwantitatieve
  Methoden}, 19, 89--104.

\bibitem[\protect\citeauthoryear{Inoue and Shirai}{Inoue and
  Shirai}{2023}]{inoue-shirai}
\textsc{Inoue, Y. and K.~Shirai} (2023): \enquote{On the Observable
  Restrictions of Limited Consideration Models: Theory and Application,}
  \emph{Economic Theory}, 75, 695--715.

\bibitem[\protect\citeauthoryear{Kalai, Rubinstein, and Spiegler}{Kalai
  et~al.}{2002}]{kalai-rubinstein-spiegler02}
\textsc{Kalai, G., A.~Rubinstein, and R.~Spiegler} (2002):
  \enquote{Rationalizing Choice Functions by Multiple Rationales,}
  \emph{Econometrica}, 70, 2481--2488.

\bibitem[\protect\citeauthoryear{Kreps}{Kreps}{1990}]{kreps90}
\textsc{Kreps, D.~M.} (1990): \emph{A Course in Microeconomic Theory},
  Princeton: Princeton University Press.

\bibitem[\protect\citeauthoryear{Kreps}{Kreps}{2012}]{kreps12}
---\hspace{-.1pt}---\hspace{-.1pt}--- (2012): \emph{Microeconomic Foundations
  I: Choice and Competitive Markets}, Princeton: Princeton University Press.

\bibitem[\protect\citeauthoryear{Liang}{Liang}{2026}]{liang-JLE}
\textsc{Liang, A.} (2026): \enquote{Using Machine Learning to Generate,
  Clarify, and Improve Economic Models,} \emph{Journal of Economic Literature},
  forthcoming.

\bibitem[\protect\citeauthoryear{Lleras, Masatlioglu, Nakajima, and
  Ozbay}{Lleras et~al.}{2017}]{lleras-masatlioglu-nakajima-ozbay}
\textsc{Lleras, J.~S., Y.~Masatlioglu, D.~Nakajima, and E.~Y. Ozbay} (2017):
  \enquote{When More is Less: Limited Consideration,} \emph{Journal of Economic
  Theory}, 170, 70--85.

\bibitem[\protect\citeauthoryear{Mackworth}{Mackworth}{1977}]{mackworth77}
\textsc{Mackworth, A.~K.} (1977): \enquote{Consistency in Networks of
  Relations,} \emph{Artificial Intelligence}, 8, 99--118.

\bibitem[\protect\citeauthoryear{Manzini}{Manzini}{2023}]{manzini23}
\textsc{Manzini, P.} (2023): \enquote{Mixture Choice Function Datasets,}
  \emph{author's personal website}.

\bibitem[\protect\citeauthoryear{Manzini and Mariotti}{Manzini and
  Mariotti}{2007}]{manzini-mariotti07}
\textsc{Manzini, P. and M.~Mariotti} (2007): \enquote{Sequentially
  Rationalizable Choice,} \emph{American Economic Review}, 97, 1824--1839.

\bibitem[\protect\citeauthoryear{Manzini and Mariotti}{Manzini and
  Mariotti}{2014}]{manzini-mariotti14}
---\hspace{-.1pt}---\hspace{-.1pt}--- (2014): \enquote{Welfare Economics and
  Bounded Rationality: The Case for Model-Based Approaches,} \emph{Journal of
  Economic Methodology}, 21, 343--360.

\bibitem[\protect\citeauthoryear{Mas-Colell, Whinston, and Green}{Mas-Colell
  et~al.}{1995}]{mwg}
\textsc{Mas-Colell, A., M.~D. Whinston, and J.~R. Green} (1995):
  \emph{Microeconomic Theory}, New York: Oxford University Press.

\bibitem[\protect\citeauthoryear{Masatlioglu, Nakajima, and Ozbay}{Masatlioglu
  et~al.}{2012}]{masatlioglu-nakajima-ozbay12}
\textsc{Masatlioglu, Y., D.~Nakajima, and E.~Y. Ozbay} (2012):
  \enquote{Revealed Attention,} \emph{American Economic Review}, 102,
  2183--2205.

\bibitem[\protect\citeauthoryear{Masatlioglu and Ok}{Masatlioglu and
  Ok}{2005}]{masatlioglu-ok05}
\textsc{Masatlioglu, Y. and E.~A. Ok} (2005): \enquote{Rational Choice with
  Status Quo Bias,} \emph{Journal of Economic Theory}, 121, 1--29.

\bibitem[\protect\citeauthoryear{Nightingale, \"{O}zg\"{u}r Akg\"{u}n, Gent,
  Jefferson, and Miguel}{Nightingale et~al.}{2014}]{SavileRow}
\textsc{Nightingale, P., \"{O}zg\"{u}r Akg\"{u}n, I.~P. Gent, C.~Jefferson, and
  I.~Miguel} (2014): \enquote{Automatically Improving Constraint Models in
  {S}avile {R}ow through Associative-Commutative Common Subexpression
  Elimination,} in \emph{Proceedings of the 20th International Conference on
  Principles and Practice of Constraint Programming (CP 2014)}, Springer,
  590--605.

\bibitem[\protect\citeauthoryear{Ok, Ortoleva, and Riella}{Ok
  et~al.}{2015}]{ok-ortoleva-riella15}
\textsc{Ok, E.~A., P.~Ortoleva, and G.~Riella} (2015): \enquote{Revealed
  (P)reference Theory,} \emph{American Economic Review}, 105, 299--321.

\bibitem[\protect\citeauthoryear{R{\'e}gin}{R{\'e}gin}{2005}]{regin04}
\textsc{R{\'e}gin, J.-C.} (2005): \enquote{Global Constraints and Filtering
  Algorithms,} in \emph{Global Constraints}, Wiley, preprint available at
  \url{https://www.constraint-programming.com/people/regin/papers/global.pdf}.

\bibitem[\protect\citeauthoryear{Richter}{Richter}{2020}]{richter20}
\textsc{Richter, M.} (2020): \enquote{Choice Theory via Equivalence,}
  \emph{Working Paper}.

\bibitem[\protect\citeauthoryear{Rossi, van Beek, and Walsh}{Rossi
  et~al.}{2006}]{rossi-vanbeek-walsh06}
\textsc{Rossi, F., P.~van Beek, and T.~Walsh} (2006): \enquote{Introduction,}
  in \emph{Handbook of Constraint Programming}, ed. by F.~Rossi, P.~van Beek,
  and T.~Walsh, Elsevier, chap.~1.

\bibitem[\protect\citeauthoryear{Rubinstein}{Rubinstein}{2006}]{rubinstein06}
\textsc{Rubinstein, A.} (2006): \emph{Lecture Notes in Microeconomic Theory},
  Princeton, NJ: Princeton University Press.

\bibitem[\protect\citeauthoryear{Rubinstein and Salant}{Rubinstein and
  Salant}{2008}]{rubinstein_salant_hndbk}
\textsc{Rubinstein, A. and Y.~Salant} (2008): \enquote{Some Thoughts on the
  Principle of Revealed Preference,} in \emph{The Foundations of Positive and
  Normative Economics: A Handbook}, ed. by A.~Caplin and A.~Schotter, New York:
  Oxford University Press, 115--124.

\bibitem[\protect\citeauthoryear{Rubinstein and Salant}{Rubinstein and
  Salant}{2012}]{rubinstein-salant12}
---\hspace{-.1pt}---\hspace{-.1pt}--- (2012): \enquote{Eliciting Welfare
  Preferences from Behavioural Data Sets,} \emph{Review of Economic Studies},
  79, 375--387.

\bibitem[\protect\citeauthoryear{Salant and Rubinstein}{Salant and
  Rubinstein}{2008}]{salant-rubinstein08}
\textsc{Salant, Y. and A.~Rubinstein} (2008): \enquote{(\textit{{A}},
  \textit{f}): Choice with Frames,} \emph{Review of Economic Studies}, 75,
  1287--1296.

\bibitem[\protect\citeauthoryear{Selten}{Selten}{1991}]{selten91}
\textsc{Selten, R.} (1991): \enquote{Properties of a Measure of Predictive
  Success,} \emph{Mathematical Social Sciences}, 21, 153--167.

\bibitem[\protect\citeauthoryear{Smeulders, Spieksma, Cherchye, and De{
  }Rock}{Smeulders et~al.}{2014}]{smeulders-spieksma-cherchye-derock}
\textsc{Smeulders, B., F.~C.~R. Spieksma, L.~Cherchye, and B.~De{ }Rock}
  (2014): \enquote{Goodness-of-Fit Measures for Revealed Preference Tests:
  Complexity Results and Algorithms,} \emph{{ACM} Transactions on Economics and
  Computation}, 2, 3:1--3:16.

\bibitem[\protect\citeauthoryear{van Beek}{van Beek}{2006}]{vanbeek06}
\textsc{van Beek, P.} (2006): \enquote{Backtracking Search Algorithms,} in
  \emph{Handbook of Constraint Programming}, ed. by F.~Rossi, P.~van Beek, and
  T.~Walsh, Elsevier, chap.~4.

\end{thebibliography}

\pagebreak

\appendix

\section*{\centering \LARGE Appendix}

\section{All compatible primitives in the Example\label{subsection:enumerated-instances}}

\paragraph{RS solutions:} 

$$ 
\begin{array}{llllll}
	1. & y\succ_1 z & \& & z\succ_2 x,\, x\succ_2 y &&  
	\text{(\textbf{RS}$_{acyc}\setminus$\textbf{RS}$_{order}$)} 
	\\
	2. & y\succ_1 z & \& & z\succ_2 x,\, x\succ_2 y,\, z\succ_2 y && 
	\text{(\textbf{RS}$_{order}$)} 
	\\
	3. & y\succ_1 z & \& & z\succ_2 x,\, x\succ_2 y,\, y\succ_2 z && 
	\text{(\textbf{RS}$_{asym}\setminus$\textbf{RS}$_{acyc}$)} 
	\\
	4. & x\succ_1 y,\, y\succ_1 z & \& & z\succ_2 x && 
	\text{(\textbf{RS}$_{acyc}\setminus$\textbf{RS}$_{order}$)} 
	\\
	5. & x\succ_1 y,\, y\succ_1 z & \& & z\succ_2 x,\, z\succ_2 y && 
	\text{(\textbf{RS}$_{acyc}\setminus$\textbf{RS}$_{order}$)} 
	\\
	6. & x\succ_1 y,\, y\succ_1 z & \& & y\succ_2 z,\, z\succ_2 x && 
	\text{(\textbf{RS}$_{acyc}\setminus$\textbf{RS}$_{order}$)} 
	\\
	7. & x\succ_1 y,\, y\succ_1 z & \& & z\succ_2 x,\, y\succ_2 x && 
	\text{(\textbf{RS}$_{acyc}\setminus$\textbf{RS}$_{order}$)} 
	\\
	8. & x\succ_1 y,\, y\succ_1 z & \& & z\succ_2 y,\, y\succ_2 x,\, z\succ_2 x && 
	\text{(\textbf{RS}$_{acyc}\setminus$\textbf{RS}$_{order}$)} 
	\\
	9.& x\succ_1 y,\, y\succ_1 z & \& & y\succ_2 z,\, z\succ_2 x,\, y\succ_2 x && 
	\text{(\textbf{RS}$_{acyc}\setminus$\textbf{RS}$_{order}$)} 
	\\
	10.& x\succ_1 y,\, y\succ_1 z & \& & z\succ_2 x,\, x\succ_2 y && 
	\text{(\textbf{RS}$_{acyc}\setminus$\textbf{RS}$_{order}$)} 
	\\
	11.& x\succ_1 y,\, y\succ_1 z & \& & z\succ_2 x,\, x\succ_2 y,\, z\succ_2 y && 
	\text{(\textbf{RS}$_{acyc}\setminus$\textbf{RS}$_{order}$)} 
	\\
	12.& x\succ_1 y,\, y\succ_1 z & \& & z\succ_2 x,\, x\succ_2 y,\, y\succ_2 z && 
	\text{(\textbf{RS}$_{asym}\setminus$\textbf{RS}$_{acyc}$)} 
\end{array}
$$

\paragraph{CLA solutions:} 

$$
\begin{array}{llll}
	1. & x\succ y \succ z { } & \& & { } 
	(A_1,A_2,A_3,A_4) \xrightarrow[]{\Gamma} 
	(\{x,y\}, \{y,z\}, \{z\}, \{x,y\})
	\\
	2. & x\succ y \succ z { } & \& & { } 
	(A_1,A_2,A_3,A_4) \xrightarrow[]{\Gamma} 
	(\{x,y\}, \{y\}, \{z\}, \{x,y\})
	\\
	3. & x\succ y \succ z { } & \& & { } 
	(A_1,A_2,A_3,A_4) \xrightarrow[]{\Gamma} 
	(\{x\}, \{y\}, \{z\}, \{x,y,z\})
	\\
	4. & x\succ y \succ z { } & \& & { } 
	(A_1,A_2,A_3,A_4) \xrightarrow[]{\Gamma} 
	(\{x\}, \{y,z\}, \{z\}, \{x,y,z\})
	\\
	5. & x\succ y \succ z { } & \& & { } 
	(A_1,A_2,A_3,A_4) \xrightarrow[]{\Gamma} 
	(\{x,y\}, \{y\}, \{z\}, \{x,y,z\})
	\\
	6. & x\succ y \succ z { } & \& & { } 
	(A_1,A_2,A_3,A_4) \xrightarrow[]{\Gamma} 
	(\{x,y\},\{y,z\},\{z\},\{x,y,z\})
	\\
	7. & x\succ z \succ y { } & \& & { } 
	(A_1,A_2,A_3,A_4) \xrightarrow[]{\Gamma} 
	(\{x,y\},\{y\},\{z\},\{x,y\})
	\\
	8. & x\succ z \succ y { } & \& & { } 
	(A_1,A_2,A_3,A_4) \xrightarrow[]{\Gamma} 
	(\{x\}, \{y\}, \{z\}, \{x,y,z\})
	\\
	9. & x\succ z \succ y { } & \& & { } 
	(A_1,A_2,A_3,A_4) \xrightarrow[]{\Gamma} 
	(\{x,y\}, \{y\}, \{z\}, \{x,y,z\})
	\\
	10. & z\succ x \succ y { } & \& & { } 
	(A_1,A_2,A_3,A_4) \xrightarrow[]{\Gamma} 
	(\{x,y\},\{y\},\{z\},\{x,y\})
	\\
	11. & z\succ x \succ y { } & \& & { } 
	(A_1,A_2,A_3,A_4) \xrightarrow[]{\Gamma} 
	(\{x,y\}, \{y\}, \{x,z\}, \{x,y\}) 
	\\
\end{array}
$$

\paragraph{OC solutions:} 

$$
\begin{array}{llll}
	1. & y\succ z \succ x { } & \& & { } 
	(A_1,A_2,A_3,A_4) \xrightarrow[]{\Gamma} 
	(\{x\},\{y\},\{x,z\},\{x\})
	\\
	2. & y\succ z \succ x { } & \& & { } 
	(A_1,A_2,A_3,A_4) \xrightarrow[]{\Gamma} 
	(\{x\},\{y,z\},\{x,z\},\{x\})
	\\
	3. & z\succ y \succ x { } & \& & { } 
	(A_1,A_2,A_3,A_4) \xrightarrow[]{\Gamma} 
	(\{x\},\{y\},\{x,z\},\{x\})
	\\
	4. & z\succ x \succ y { } & \& & { } 
	(A_1,A_2,A_3,A_4) \xrightarrow[]{\Gamma} 
	(\{x\},\{y\},\{x,z\},\{x\})
	\\
	5. & z\succ x \succ y { } & \& & { } 
	(A_1,A_2,A_3,A_4) \xrightarrow[]{\Gamma} 
	(\{x,y\},\{y\},\{x,z\},\{x\})
	\\
	6. & z \succ x \succ y { } & \& & { } 
	(A_1,A_2,A_3,A_4) \xrightarrow[]{\Gamma} 
	(\{x,y\},\{y\},\{x,z\},\{x,y\})
\end{array}
$$
\hfill $\blacklozenge$

\section{All \textit{undominated} model primitives in the Exam\nolinebreak ple}

Applying the dominance notions of Definitions 
\ref{dfn:dominance-rs} and \ref{dfn:dominance-la}
to the list of perfectly compatible instances that are 
enumerated in Section \ref{subsection:enumerated-instances}
in relation to the example choices in \eqref{eq:example-data} 
leads to the following refined elicitations per model or class thereof 
(the \# numberings follow the respective list orders of that section):

\begin{table}[!htbp]
\begin{tabular}{lccc}
\textbf{RS$_{order}^{undominated}$ :} 
& \#2 
& 
& 
\\
& 
&
&
\\
&&&\\
\textbf{RS$_{acyc}^{undominated}$ :} 
& \#8 
& \#9 
& \#11 
\\
& (dominates \#5,7) 
& (dominates \#6)
& (dominates \#1,4,10)
\\
&&&\\
\textbf{RS$_{asym}^{undominated}$ :} 
& \#12 
&
&
\\
& (dominates \#3)
&
&
\\
&&&\\
\textbf{CLA$^{undominated}$ :} 
& \#6 
& \#9 
& \#11 
\\
& (dominates \#1--5) 
& (dominates \#7,8) 
& (dominates \#10)
\\
&&&\\
\textbf{OC$^{undominated}$ :} 
& \#2 
& \#3 
& \#6 
\\
& (dominates \#1)
&  
& (dominates \#4,5)
\end{tabular}
\end{table}

\hfill $\blacklozenge$\ \\

\section{Executed CP models and example dataset}

\section*{Essence encoding of a choice dataset}

\parindent=0pt
\parskip=0pt

Saved as a \texttt{.param} file. Comments preceded by \$. Indentation for illustration only.\\

\texttt{letting dataset be sequence$($}

\texttt{$(\{2,4,3,1\}, 4)$,} \qquad \$ item `4' is chosen at menu `$\{2,4,3,1\}$' 

\texttt{$(\{2,4,1\}, 4)$,}

\texttt{$(\{2,3,1\}, 3)$,} 

\texttt{$(\{3,1,4\}, 4)$,}

\texttt{$(\{4,3,2\}, 3)$,}

\texttt{$(\{1,2\}, 1)$,}

\texttt{$(\{4,1\}, 4)$,}

\texttt{$(\{4,2\}, 4)$,}

\texttt{$(\{3,2\}, 3)$,}

\texttt{$(\{1,3\}, 3)$,}

\texttt{$(\{3,4\}, 4))$}

\section*{Essence model for Rational Shortlisting}

Saved as an \texttt{.essence} file. Comments preceded by \$. Indentation for illustration only.\\

\texttt{given dataset : sequence of (set of int, int)}

\texttt{letting n be max$([$ x $|$ $(\_,$ $($xs, \_$))$ $<$- dataset, x $<$- xs $])$}\\
\texttt{letting options be domain int(1..n)}

\texttt{letting nbObservations be $|$dataset$|$}

\texttt{find omitList : sequence (size nbObservations) of bool}\\

\$ (i, j) means i is preferred to j by asymmetric relation \texttt{rel1}\\
\texttt{find rel1 : relation (minSize 1, aSymmetric) of (options * options)}\\

\$ For each entry in the dataset, we will record the survivors after applying \texttt{rel1}.\\
\$ adding \texttt{minSize 1} to the set here disallows Rational Choice solutions\\
\texttt{find survivor1 : sequence (size nbObservations) of set of options}

\texttt{find rel2 : relation (minSize 1, aSymmetric) of (options * options)}\\

\$ For each entry in the \texttt{survivor1} set, we will record the survivors after applying \texttt{rel2}.\\
\$ size 1 - single valued choice\\
\texttt{find survivor2 : sequence (size nbObservations) of set (size 1) of options}

\texttt{such that}

\qquad \texttt{forAll row : int(1..nbObservations) .}

\qquad\qquad \texttt{survivor1(row) subsetEq dataset(row)[1] $\slash\backslash$}

\qquad\qquad \texttt{((omitList(row) $\slash\backslash$ !(dataset(row)[2] in survivor1(row))) $\backslash\slash$}

\qquad\qquad\texttt{(!omitList(row) $\slash\backslash$ (dataset(row)[2] in survivor1(row))))  $\slash\backslash$}

\qquad\qquad\$ Element survives if it is not dominated by any other element of the set.

\qquad\qquad\texttt{forAll x in dataset(row)[1] .}

\qquad\qquad\texttt{x in survivor1(row) $<$-$>$ forAll y in dataset(row)[1] . !rel1(y,\nolinebreak x)}\\

\texttt{such that}

\qquad \texttt{forAll row : int(1..nbObservations) .}

\qquad\qquad \texttt{survivor2(row) subsetEq survivor1(row) $\slash\backslash$}

\qquad\qquad \texttt{((omitList(row) $\slash\backslash$ !(dataset(row)[2] in survivor2(row))) $\backslash\slash$}

\qquad\qquad \texttt{(!omitList(row) $\slash\backslash$ (dataset(row)[2] in survivor2(row)))) $\slash\backslash$}

\qquad\qquad \texttt{forAll x in survivor1(row) .}

\qquad\qquad \texttt{x in survivor2(row) $<$-$>$ forAll y in survivor1(row) . !rel2(y,\nolinebreak x)}\\

\$ Distance score is min number of elements to 
be removed from dataset\\ 
\$ to comply perfectly with the model.

\texttt{find score : int(0..nbObservations)}

\texttt{such that score = sum row : int(1..nbObservations) . toInt(omitList(row))}\\

\$ Unique solutions are defined by the preference relations, not by survivor sets.

\texttt{branching on [rel1, rel2]}

\pagebreak

\section*{Essence model of Choice with Limited Attention}

Saved as an \texttt{.essence} file. Comments preceded by \$. Indentation for illustration only.\\

\texttt{given dataset : sequence of (set of int, int)}

\texttt{letting n be max$([$ x $|$ (\_, (xs, \_)) $<$- dataset, x $<$- xs $])$}\\
\texttt{letting options be domain int(1..n)}\\

\texttt{letting nbObservations be |dataset|}\\

\texttt{find omitList : sequence (size nbObservations) of bool}\\

\$ Maximal size to find all complete solutions.\\
\texttt{find G : function (total, size 2**$|$options$|$-1)
set (minSize 1) of options --$>$ set (minSize 1) of options}\\

\$ (i, j) means i is preferred to j under strict linear order \texttt{rel}\\
\texttt{find rel : relation (aSymmetric, connex, transitive) of (options * options)}\\

\$ $\Gamma(A_i)$ is a non-empty subset of $A_i$\\
\texttt{such that forAll (menu, filter) in G .}

\qquad \texttt{filter subsetEq menu $\slash\backslash$ 
	$|$filter$|$ $>$ 0 $\slash\backslash$ $|$menu$|$ $>$ 0}\\

\$ If an element $x$ is not in $\Gamma(A_i)$ then $\Gamma(A_i) = \Gamma(A_i \setminus\{x\})$

\texttt{such that forAll row : int(1..nbObservations) .}

\qquad \texttt{forAll x in dataset(row)[1] .}

\qquad\qquad \texttt{!(x in G(dataset(row)[1])) -$>$ 
(G(dataset(row)[1]) =}
 
\qquad\qquad \texttt{G((dataset(row)[1] - {x})))}\\

\$ Choice, $x$, is selected from $\Gamma(A_i)$ 
where no other element of $\Gamma(A_i) \succ x$\\
\texttt{such that forAll row : int(1..nbObservations) .}

\qquad \texttt{((omitList(row) $\slash\backslash$ !(dataset(row)[2] 
	in G(dataset(row)[1]))) $\backslash\slash$}
	
\qquad \texttt{(!omitList(row) $\slash\backslash$ (dataset(row)[2] in G(dataset(row)[1])))) $\slash\backslash$}

\qquad \texttt{(forAll (i, j) in rel . (i in G(dataset(row)[1]) $\slash\backslash$ 
j in G(dataset(row)[1]))}

\qquad \texttt{-$>$ dataset(row)[2] != j)}\\

\texttt{find score : int(0..nbObservations)}\\
\texttt{such that score = sum row : int(1..nbObservations) . toInt(omitList(row))}

\pagebreak

\section*{Essence model for Overwhelming Choice}

Saved as an \texttt{.essence} file. Comments preceded by \$. Indentation for illustration only.\\

\texttt{given dataset : sequence of (set of int, int)}\\
\texttt{letting n be max$([$ x $|$ (\_, (xs, \_)) $<$- dataset, x $<$- xs $])$}\\
\texttt{letting options be domain int(1..n)}\\
\texttt{letting nbObservations be $|$dataset$|$}\\
\texttt{find omitList : sequence (size nbObservations) of bool}\\
\texttt{find G : function (total, size 2**$|$options$|$-1) 
set (minSize 1) of options --$>$ set (minSize 1) of options}\\

\$ (i, j) means i is preferred to j under strict linear order \texttt{rel}\\
\texttt{find rel : relation (aSymmetric, connex, transitive) of (options * options)}\\

\$ $\Gamma(A_i)$ is a non-empty subset of $A_i$\\ 
\texttt{such that forAll (menu, filter) in G .}

\qquad \texttt{filter subsetEq menu $\slash\backslash$ $|$filter$|$ $>$ 0 $\slash\backslash$ $|$menu$|$ $>$ 0}\\

\$ Choice, $x$,  is selected from $\Gamma(A_i)$ where no other element of $\Gamma(A_i) \succ x$\\
\texttt{such that forAll row : int(1..nbObservations) .}

\qquad \texttt{((omitList(row) $\slash\backslash$ !(dataset(row)[2] in G(dataset(row)[1]))) $\backslash\slash$} 

\qquad \texttt{(!omitList(row) $\slash\backslash$ (dataset(row)[2] in G(dataset(row)[1])))) $\slash\backslash$}

\qquad \texttt{(forAll (i, j) in rel . (i in G(dataset(row)[1]) $\slash\backslash$ 
j in G(dataset(row)[1]))} 

\qquad \texttt{-$>$ dataset(row)[2] != j)}\\

\$ Extra structure constraint: if $S \subset T$, then $(\Gamma(T) \cap S) \subseteq \Gamma(S)$\\
\texttt{such that forAll (menuT, filterT) in G .}

\qquad 
\texttt{forAll (menuS, filterS) in G .}

\qquad \qquad 
\texttt{menuS subset menuT -$>$ (filterT intersect menuS) subsetEq filterS}\\

\texttt{find score : int(0..nbObservations)}

\texttt{such that score = sum row : int(1..nbObservations) . toInt(omitList(row))}

\end{document}